\documentclass[runningheads]{llncs}
\usepackage[T1]{fontenc}

\usepackage[top=2cm, bottom=2cm, right=2cm, left=2cm]{geometry}
\usepackage[english]{babel}
\usepackage{enumerate}
\usepackage{graphicx}
\usepackage{svg}
\usepackage{listings}
\usepackage{amsmath}
\usepackage{amssymb}
\usepackage{listings}
\usepackage{tikz}
\usetikzlibrary{positioning}
\usepackage{textcomp}
\usepackage{stackengine}
\usepackage[capitalize,noabbrev,nameinlink]{cleveref}
\usepackage[title]{appendix}
\usepackage{array}
\usepackage{multicol}
\usepackage{fancybox}
\usepackage{lmodern}

\usepackage{fancyvrb}

\usepackage{listings}
\begin{document}

%
%

%
\title{RDF Surfaces: Enabling Classical Negation on the Semantic Web}
%
%
\author{Patrick Hochstenbach\inst{1,2}\orcidID{0000-0001-8390-6171} \and
Mathijs van Noort\inst{2}\orcidID{0000-0003-2719-5402} \and
Dörthe Arndt\inst{3}\orcidID{0000-0002-7401-8487} \and
Rebekka Martens\inst{3} \and
Jos De Roo\inst{2}\orcidID{0000-0001-8862-0666} \and
Ruben Verborgh\inst{2}\orcidID{0000-0002-8596-222X} \and
Pieter Bonte\inst{4}\orcidID{0000-0002-8931-8343} \and
Femke Ongenae\inst{2}\orcidID{0000-0003-2529-5477}}
\authorrunning{P. Hochstenbach et al.}
%
\institute{Ghent University Library, Ghent University, Belgium \and
IDLab, Ghent University - imec, Belgium \and
International Center for Computational Logic, Technische Universität Dresden, Germany \and
Department of Computer Science, KU Leuven Campus Kulak, Belgium \\
\email{patrick.hochstenbach@ugent.be}}
\maketitle              
\begin{abstract}
The Resource Description Framework (RDF) is a fundamental technology in the Semantic Web, enabling the representation and interchange of structured data. However, RDF lacks the capability to express negated statements in a generic way. As a result, exchanging negative information on a Web scale is thus far restricted to specific cases and predefined statements. The ability to negate (virtually) any RDF statement allows for a comprehensive way to refute, deny or otherwise invalidate claims on a Web scale. Via an intermediate step of a diagrammatic approach to logical expressions called Peirce graphs, we introduce RDF Surfaces, an extension of RDF that incorporates the concept of classic negation, known from first-order logic. 
Overall, RDF Surfaces provides an abstract, visual approach to negation within the Semantic Web, offering a more general and widely applicable approach than previous attempts at incorporating negation. Aside from a (traditional) programmatic syntax, RDF Surfaces can also be represented visually by means of diagrams inspired by Peirce graphs.  We demonstrate negation via RDF Surfaces and how to reason upon it in illustrative use cases drawn from the domains of academic publishing and eHealth. We hope this vision paper attracts new implementers and opens the discussion to its formal specification.

\end{abstract}
%
%
%

\section{Introduction}\label{s1}

Reasoning with classical negation has attracted interest in the Semantic Web since its early days. Classical negation has properties similar to processing boolean values with the \texttt{NOT} operator in computer languages. Classical negation can take "true" to "false" and vice versa. It follows the law of the excluded middle: any statement $P$ is either "true" or "false". If some $P$ is "true", then a double negated $P$ is also "true" (and vice versa for "false"). This form of negation is also explicit: "false" is not a default value; it cannot be assumed; it should be stated. All other forms of negation that do not follow these principles -- there are many variants --  are "weaker" forms of negation. The interest in "strong" classical negation can be attributed to a desire to fulfill one of its core design principles. Just as the Web is open-ended, allowing anyone to create a webpage on a server and link to any other webpage, the Semantic Web aspires to enable anyone to express any statement about any topic~\cite{koivunen_w3c_semweb}. The Resource Description Framework (RDF)~\cite{RDF} is the W3C recommendation that defines the language to express statements about anything in the universe in the form of RDF triples. Using RDF, not only Web resources can be described, but physical objects, abstract concepts, and, in general, anything that can be given an identifier. However, there are also limitations. RDF lacks the ability to express classical "strong" negation, explicitly stating negative information that also follows the law of the excluded middle and all other classical negation properties. Additionally, RDF can express existential quantification (using blank nodes): statements about \textit{one} or more resources, but lacks the expressivity to create statements about \textit{zero} or more resources: universal quantification. As we see below, there are reasons to desire such features, but there are also some compelling reasons why classical negation has mostly been avoided.

\subsection{Why is classical negation and universal quantification desirable?}\label{s1.1}

We provide three argument for classical negation and universal quantification.

First, negation and universal quantification are applicable across various use cases. Straccia and Casini~\cite{straccia_minimal_2022} describe how, in medicine, it is important to distinguish between the absence of biochemical reactions between substances and not knowing about their existence, which results in a need for explicitly stating negative information. Wagner~\cite{wagner_towards_2005} makes the case for a generalized Web logic based on classic negation and quantification to drive business processes. In an earlier paper, Wagner~\cite{goos_web_2003} demonstrated that two types of negation are necessary to interpret data effectively. The monotonic strong (classical) negation "Patient X did not take pill-B" expresses negative knowledge. The non-monotonic weak negation "Patient X is not registered at a hospital" expresses a negation as failure (NAF). Esteves~\cite{esteves_odrl_2021} and Kebede~\cite{kebede_critical_2021} advocate for the use of negation in making policy information on permissions, prohibitions, and obligations enforceable through the Open Digital Rights Language (ODRL)~\cite{ODRL}. They also suggest that negation can facilitate conflict resolution while creating policy documents.

Second, democratic and social reasons can be provided to desire classical negation on the Semantic Web. We live in times of massive information flows trying to influence users' social and political worldviews. Deciding what resources can be trusted, or even evaluating contrasting views, is challenging for many Web citizens. The Semantic Web is not the harbinger of absolute truth: expressing a statement as RDF does not automatically make it closer to absolute truth. Every application that processes information on the Semantic Web relies on a lesser form of truth, a relative truth: trust. By "asserting an RDF triple," a Semantic Web application assumes a triple as being "true" (whatever that means in the real world).\footnote{In this paper we will quote "true" and "false" to remind the reader about this relative truth. However, with important logical consequences.} Combining multiple RDF triples forms those triples' logical conjunction (AND). Based on this relative truth of a set of triples, boolean queries can be executed using the SPARQL Query language~\cite{sparql}, and new RDF triples can be inferred using the RDF Schema language~\cite{brickley_rdf_2014}.  However, without classical negation, explicitly stating what is not the case, there is no explicit notion of contradiction on the Semantic Web. This limits the applicability of formal ways to validate statements and discover contrasting views on the Semantic Web.

Third, there could be scientific reasons to add these features to the Semantic Web. Classical negation combined with existential quantification in RDF, as we will see later in the paper, provides the expressivity of first-order logic (FOL), including universal quantification. The properties of this logic have been studied over centuries. Adding classical negation to the Web, even for its own sake, would give us broad opportunities to express human reasoning and scientific inquiries. It offers the chance to present facts on the Web and the logic behind them, enhancing the scientific process's transparency.

In addition to practical reasons related to solving real-world use cases, there are also technical reasons why a classical negation and, in extension, FOL have desirable properties.

FOL retains the monotonicity property. When Web logic is monotonic, adding new asserting RDF triples to a previous set of asserted RDF triples will never invalidate previous queries or inferences. What was a valid inference using an older set of RDF triples must also be a valid inference in the updated set of RDF triples. Following Hayes (2001), Web reasoning is inherently open-ended, i.e.~one can never assume all the facts about a topic are available. By consulting additional resources, new information might arise. Non-monotonic reasoning, such as NAF, assumes that information about any topic is complete. Missing information is assumed to be "false". However, this is unsafe in an open world, such as the Web, where one does not have the license to assume a statement is "false" without explicitly stating it to be "false". This is not the case for classical negation, which is explicit in the triples that should be considered "false". 

There are also grounds to believe that contradictions on the Semantic Web are inherent, not detrimental as suggested by our second point, but rather benign. Hayes refers to this as the 'diamond of confusion'~\cite{hayes_2021}: we may agree on a shared reality (whether it be absolute or relative truth) and a common method to create statements and express logical reasoning about this reality, yet still end up with contradictory conceptualizations of this reality. Hayes provides an example of how concepts can be conceptualized with and without a temporal dimension, which leads to contradictory results. For instance, a patient can be a fixed "thing" in one formalization with a name, address, and social security number; the same patient can be an "event in time" in another formalization where the patient with a fever is not the same patient without fever after giving a medication (a "thing" cannot both have a fever and not a fever). Understanding and identifying these contradictions among RDF resources is crucial for interpreting resources in querying and reasoning scenarios.

In some way, negation is already implicitly available in the Semantic Web when asserting triples. In logic, there are no alternative versions of "true". If, within a context, some RDF triple is regarded as being "true", this means that the negation of that triple is "false" in that context (regardless of the absolute truth or availability of other RDF triples on the open Web). Or, stated differently, within a context, any assertion negates the negated triple because $P \equiv \lnot \lnot P$. 

\subsection{Why was classical negation not added to RDF?}\label{s1.2}

Considering all the reasons, why can't we incorporate this form of classic negation into RDF?

Monotonic logic was welcomed by many as a desirable feature on the Web, but the full expressivity of monotonic logic in the form of FOL was not. The designers of RDF deliberately chose to exclude these features, citing Lassila~\cite{lassila_web_1998}, who expressed concern that such complex features "might discourage the acceptance and adoption of RDF within the Web community." 

A more compelling argument against logics with the expressivity of FOL is that they have been proven to be undecidable for finding all valid inferences from a knowledge base~\cite{trakhtenbrot_recursive_1953}. For machines, undecidable means that processing any arbitrary RDF with full FOL expressivity, \textit{and} finding all correct inferences, \textit{and} doing all of this in a finite amount of time is impossible. At most, two of these features can be achieved~\cite{oregan_computability_2023}. 

Finding all valid inferences in a finite time (two of the features of the previous point) is one of the core design issues in the Web Ontology Language (OWL)~\cite{OWL}. OWL2 DL is the flavour of OWL based on Description Logics (DL). OWL2 DL chose to use only those fragments of FOL that are decidable and (a second design choice) can be safely processed by machines in isolation. A safe execution makes computational reasoning tasks tractable (executable in polynomial time). However, two compromises need to be made to achieve this form of computability: negation and quantification need to be introduced in a weakened form, and OWL2 DL and its sublanguages need to abandon the semantics of RDF when expressing Web logic. OWL2 DL introduces "Direct Semantics"~\cite{motik_owl_2012} as an alternative for "RDF Semantics." Only the former has the desired decidable and tractable features.\footnote{\url{https://www.w3.org/TR/2012/REC-owl2-rdf-based-semantics-20121211/\#Example\_on\_Semantic\_Differences}} However, Direct Semantics disregards the RDF nature of the OWL2 DL formulas in its semantics.

\subsection{Rationale for proposing the addition of classical negation and FOL expressivity to RDF}\label{s1.3}

Our rationale for adding classic negation and FOL expressivity to RDF is not novel. They were articulated in Hayes’ “BLOGIC” invited talk at IWSC 2009~\cite{hayes_blogic_2009}. In his argumentation, RDF is portable, i.e.~any, RDF triple expresses the same data irrespective of the processing environment. However, this is not the case for current Web logics. Combinations of RDFS and OWL2 DL (and even within OWL2 DL) come with semantics that do not always agree. Web logics that use only a fraction of FOL does not commute: different fragments can disagree on what can be concluded from an RDF knowledge base. These concerns for the portability of Web logic led Hayes to propose weB LOGIC (BLOGIC) as a new approach for portable logic on the Web, rooted in the theory of \textit{existential graphs} by Charles Sanders Peirce (1839-1914)~\cite{sowa_john_f_knowledge_2000}.
In Hayes's vision, BLOGIC in the form of existential graphs mitigates the limited expressivity of RDF and the limitations on the portability of logic on the Web. To incorporate existential graphs, RDF needs to introduce two additional concepts: a \textit{surface} as a boundary for negated triples and for collections of \textit{graffiti} (blank) nodes that act as existentially quantified variables. Together with the standard assertion of triples and their conjunction, the full expressivity of FOL can be achieved without losing the structure and semantics of existing RDF resources.

Achieving BLOGIC with FOL expressivity again introduces undecidability to the  Semantic Web, which is precisely what the RDF designers aimed to avoid. We argue that any \textit{portable} Web logic in RDF will likely not be decidable. Decidability requires a delicate choice in limiting the expressive power of Web logics. While fragments of FOL may individually lead to decidability, their combination often does not, and certainly not in combination with the semantics of RDF. To communicate and transport Web logic as RDF, undecidability will be a fact we must live with and not a limiting choice. 

Undecidability, in general, is not a showstopper on the Web. The satisfiability of the RDF Query language (SPARQL) is undecidable\cite{hartig_2012}\cite{zhang_2016}. Billions of websites and Web applications exchange HTML, CSS, and JavaScript code in combination, fully Turing complete, thus providing a daily undecidable halting problem on the Web stack. Despite this, the Web has continued to evolve and thrive. Real-world use cases might not require solving the most extreme computability problems. In our paper, we will highlight two use cases that can be solved by implementing our translation of the BLOGIC vision in RDF, which we call \textit{RDF Surfaces}.

We do not assume that a single machine will solve the undecidable problem and simultaneously provide a solution that can accept arbitrary knowledge input, produce all valid inferences, and always find these results in a finite amount of time. However, one of these features can be dropped to turn an undecidable problem into a decidable one where machines can assist humans in decision-making.

Our standpoint is a bit provocative. We assume that human knowledge is, by nature, undecidable and contradictory. Even in science, only romanticized views regard it as a "flawless building" of knowledge,  only containing positive information and results~\cite{matosin_negativity_2014}. A true Semantic Web should encompass more than just information and logic that machines can process. Human knowledge is decentralized and can be contradictory. Machine intelligence, if it wants to be in any way compared to what humans can produce, needs to be decentralized and able to be contradictory too. These contradictions should not be hidden, but expressed openly in portable syntax and logic.

In summary, a Semantic Web that is closer to enabling anyone to express any statement about any topic requires:

\begin{enumerate}
    \item Expressing the classic negation of every possible statement.
    \item Providing the full expressivity of FOL (including universal quantification).
    \item Adhere to the RDF program of portability.
\end{enumerate}

\subsection{Contribution}

As far as we know, no attempt has been made to implement classical negation, as proposed by the BLOGIC vision, in a concrete RDF syntax and implementation since Hayes' talk in 2009. Our paper will introduce RDF Surfaces as an extension of RDF's simple interpretation based on Peirce's existential graphs, with the extended semantics of classical negation and the expressivity of FOL. In this vision paper, we aim to translate Hayes' BLOGIC vision into a concrete RDF syntax, investigate the expressivity of its semantics, test the applicability in real-world use cases, demonstrate initial implementation steps, and encourage further research in formalization and potential implementations.

Our main contributions in this paper are thus as follows:

\begin{enumerate}
 \item \textbf{RDF Surfaces Syntax:} We apply existential graphs to RDF in the form of RDF Surfaces and demonstrate how Hayes' BLOGIC vision can be made concrete with a serialization using a subset of the Notation3 syntax.
 \item \textbf{Investigation of expressivity:} We demonstrate how two additions to the RDF model under simple entailment semantics provide the full expressivity of FOL with explicit quantification, with additions of (a) surfaces with negated contents, and (b) collections of graffiti (blank) nodes that act as existentially quantified variables.
 \item \textbf{Showcase applicability:} We provide two use cases, one from scholarly communication and one from the healthcare domain, demonstrating the need for classic negotiation. These use cases demonstrate how positive and negative data and logic can be shared using the RDF syntax with extended semantics. Both use cases implement FOL features, such as quantification, disjunctions, and implications to share logic rules.
 \item \textbf{RDF Surfaces initial reasoner implementation:} We demonstrate how RDF Surfaces and the examples of our paper can be queried using an implementation of RDF Surfaces.
\end{enumerate}

\subsection{Paper outline}

In the remainder of this paper, \cref{s2} will highlight two scenarios that will be used to illustrate the application of FOL Web logic in scholarly communication and healthcare. In \cref{s3}, we will review related work on implementing FOL and negation on the Web. In \cref{s4}, we introduce Peirce's existential graphs. In \cref{s5}, we introduce RDF Surfaces and its syntax, which implements Hayes' BLOGIC vision. In \cref{s6}, an overview will be presented of four implementations of RDF Surfaces and the properties of the most mature version we used to test the use cases of this paper. In \cref{s7}, the RDF Surfaces will be applied to the two use cases presented in \cref{s2}. In \cref{s8}, we will discuss our study's main points and further work. Finally, our conclusions will be presented in \cref{s9}.

\section{Running examples}\label{s2}

This article will use two running examples to highlight the potential for negation on the Web. Our use cases help explain this paper's abstract concepts and put their application in the context of real-world use cases. 

\subsection{Scholarly communication}\label{s2.1}

For many researchers, choosing the right place to publish is the most contentious question in their pathway to pursue an academic degree. Researchers do not need to spend much time searching online to find numerous websites from universities, libraries, publishers, and individual researchers that present the trade-offs in various publication paths. This question can be more complex than it seems. A researcher might want to give a topic more visibility by targeting a wide (nonspecialist) academic community or the general public. Alternatively, a researcher could choose to publish new results as fast as possible to claim precedence. Some scholarly communities progress in their field by sharing these fast research results as preprints in subject repositories, such as arXiv and medRxiv. However, certain journals refuse to accept research articles previously disseminated in this manner or charge high article processing charges (APC). Institutions might prefer to accept only some types of publications for satisfactory completion of a degree, for instance, only publications from high-impact journals indexed in the Web of Science (WOS) database. The academic world could utilize library databases to share lists of journals that are explicitly stated to be part of the WOS database or explicitly state them as excluded or formally removed from coverage.

It would benefit all scholarly communication network actors to share their preferences using publicly accessible preference documents. These preferences could then provide input for smart agents to suggest new venues and provide the best advice on where to publish. An example of each actor's different types of policies is sketched below in Table~\ref{schol_table_1} and Table~\ref{schol_table_2}.

Both the researcher and department preferences contain explicit and implicit negations. A journal that explicitly states that APC costs are not charged is a researcher's X preference. The department's Y demand for a venue in WOS excludes all venues that are explicitly not indexed in WOS. Journal facts make positive and negative facts clear. These facts can be published online by the journal or publisher's homepage or provided by library databases that track journal information. In our examples, journals ABC and DEF would be researcher and departmental preferences, but journal GHI and repository JKL would not simultaneously match the researcher and departmental preferences.

\begin{table}
  \begin{tabular}{ | m{22em} | m{22em} | }
    	\textbf{Researcher X Preferences} & \textbf{Department Y Preferences} \\
	\hline
	 Prefers a subject repository, a journal that does not charge APC costs or an indexed journal in WOS. &
	 The publication venue must be indexed in WOS. \\
    \end{tabular}
    \caption{Examples of possible publication venue preferences for a researcher and a department.}
    \label{schol_table_1}
\end{table}

\begin{table}
    \begin{tabular}{ | m{10em} | m{10em} | m{10em} | m{10em} | }
     	\textbf{Journal ABC Facts} & \textbf{Journal DEF Facts} & \textbf{Journal GHI Facts} & \textbf{Repository JKL Facts} \\
        \hline
	 Indexed in WOS.  Requires APC.  & Indexed in WOS. & Not indexed in WOS. Does not require APC. & A subject repository. \\
    \end{tabular}
    \caption{Facts for hypothetical journals ABC, DEF, GHI, and a repository JKL.}
    \label{schol_table_2}
\end{table}

\begin{table}
    \begin{tabular}{|c|c|c|}
        \textbf{Medicine} & \textbf{Treated affliction} & \textbf{Exclusion criteria}\\
        \hline
        High dosage of aspirin & Fever & Aspirin Allergy, Active peptic ulcer disease \\
        Low dosage of aspirin & Acute myocardial infarction & Aspirin allergy, Active peptic ulcer disease \\
        Beta-blockers & Acute myocardial infarction & Severe asthma, Chronic obstructive pulmonary disease \\
    \end{tabular}
    \caption{Overview of Medicine treatment}
    \label{tab:medicine_treatment}
\end{table}

\subsection{Medicine Prescription}\label{s2.2}

For the second use case, we show the applicability of FOL to the healthcare domain. Determining a suitable set of medications is a complex process, taking into account a patient's symptoms, the effectiveness of medication, the patient profile, and other factors. Specifically, the presence of allergies towards one or more types of medication, as well as possible secondary afflictions inferencing with some medication,  makes for a difficult process to prescribe each patient the best suitable medication. Combining a high level of required domain knowledge and the need to perform complex reasoning upon said knowledge makes this an overall hard process to capture and automate accurately.

For a given condition, multiple possible medicines are often available to provide treatment. Here, we consider the condition of an acute myocardial infarction, with possible treatments of a low dosage of aspirin or beta-blockers. Aspirin should not be prescribed if a patient is allergic to it or suffers from active peptic ulcer disease. Likewise, beta-blockers should only be prescribed when a patient does not suffer from severe asthma or chronic obstructive pulmonary disease. Furthermore, a high aspirin dosage is known to be an effective treatment for fever, with identical exclusion criteria compared to a low aspirin dosage. The information is summarised in Table~\ref{tab:medicine_treatment}.

To arrive at a suitable medicine prescription, it is necessary to reason upon negative information. A patient should only be prescribed a given drug if it does hold that said drug is an effective treatment of a patient's condition and if all the known exclusion criteria of the drug are assured to \textit{not} hold, e.g., in case of aspirin prescription, it must hold that a patient does not have an aspirin allergy nor a peptic ulcer disease.


\section{Related Work}\label{s3}

A significant body of research on defining the requirements for Web logic with negation is available with possible extensions to FOL expressivity. \cref{tab:related_works} presents an overview of the three main requirements we seek for a Web logic, i.e.~ enabling classical negation, providing the full expressivity of FOL (including universal quantification), and building upon the RDF data \& syntax to attain portability, and compares them against the solutions discussed in the following paragraphs in more detail. It can be concluded from this table that none of the available solutions fit all the requirements, hence the motivation for RDF Surfaces. 

\begin{table}[h]
    \centering
    \begin{tabular}{|c|c|c|c|}
        \textbf{Technology} & \textbf{Classic negation} & \textbf{FOL expressivity} & \textbf{RDF data \& logic}\\
        \hline
        KIF & + & + & - \\
        Common Logic & + & + & - \\
        N3Logic & - & - & + \\
        FIPA & - & - & + \\
        OWL2 DL & + (restricted) & - & + \\
        RIF & + & +(restricted) & - \\
        SWRL & - & - & + \\
        SWSL & + & + & - \\
        De Bruijn, Tsarkov, Tammet & + & + & - \\
        TPTP & + & + & - \\
        Datalog & - & - & - \\
        ASP & + & +(subset) & - \\
        SPARQL & +(limited to filters) & +(limited to filters) & + \\
    \end{tabular}
    \caption{Overview of technologies and their support for classic negation with explicit quantification for processing RDF data using an RDF syntax.}
    \label{tab:related_works}
\end{table}

In the early 1990s, the Defense Advanced Research Projects Agency (DARPA) and other funding agencies started the development of the Knowledge Interchange Format (KIF) as a machine-readable interchange format of knowledge among disparate programs with an expressivity near equivalent to FOL predicate calculus including classical negation~\cite{KIF}. KIF's Lisp-based syntax predates the Semantic Web and was, in the 1990s, the de facto exchange format in the research community. Common Logic continued the work of KIF and has since been developed and published as the ISO standard "ISO/IEC 24707:2018" as a framework for a family of logic-based languages~\cite{CommonLogic}. Common Logic can process RDF data but the syntax relies on Lisp like S-expressions. Common Logic is highly relevant for the BLOGIC vision as it includes Piercian graphical logic in the Appendix B of the ISO standard. Our paper applies this logic to the Semantic Web using an RDF syntax, ensuring compatibility with all existing RDF resources. With RDF Surfaces, we strive not to separate data and logic. No separate syntax and semantics are required to negate a set of RDF triples.

In 2000, Berners-Lee called for developing a unifying language for classical logic as an extension, or even the modification, of the RDF model. His SWeLL proposal was imaged to ``allow any Web software to read and manipulate data published by any other Web software''~\cite{SWELL}.  For all logical relations to be expressed, the SWeLL project proposal advocated negation and explicit quantification as an extension to RDF. The work on SWeLL influenced the development of N3Logic~\cite{N3Logic} as an extension of RDF so that the same language can be used for transporting logic and data. N3Logic provides negation in the form of scoped negation as failure (SNAF), i.e.~the monotonic version of NAF. Both NAF and SNAF are logical operations to reason about information missing from a knowledge graph, but cannot be used (or in a very limited form) to express negative information that classic negation requires explicitly. Both NAF and SNAF do not have all the desired properties of classical negation. In N3Logic, SNAF can only be used as part of an implication and not as part of the data. It is possible to create a negated statements and negated graphs by setting the consequent of an implication "false" $G \rightarrow \texttt{false}$, but this negated graph does not have the properties of a classic negation. A double negated graph cannot be interpreted as "true".

In 2001, the FIPA RDF Content Language Specification~\cite{fipa} was created to specify how RDF can be used as a message content language in the communication acts of FIPA-compliant agents. It proposes a method to express negated RDF facts by adding a believe or disbelieve in the facts. The FIPA proposal adds a \texttt{fipa:Proposition} and relies on reification of RDF triples. A \texttt{fipa:believe} predicate can be added to the reified triple to express a boolean trust. Using this mechanism a single triple can be interpreted as "false", but it is not a classical negation for the same reason as N3Logic "false" is not a classical negation.

Related developments were made in 2004 with the Web Ontology Language (OWL) as an extension of RDF. The newest version, OWL 2~\cite{OWL}, has several profiles, of which OWL 2 DL is based on fragments of FOL.  In all OWL 2 DL profiles, negation is available in the form of \texttt{owl:complementOf} and \texttt{owl:NegativePropertyAssertion}, but these negations are restricted to specific cases in the profiles to prevent the halting problem and ensure the language remains decidable.
As an example, using the \texttt{owl:NegativePropertyAssertion} it is possible to state that two individual do not have the relation \texttt{:hasParent}, 

\begin{lstlisting}[numbers=none]
:John a owl:NamedIndividual .
:Mary a owl:NamedIndividual .

[ a owl:NegativeObjectPropertyAssertion ;
  owl:sourceIndividual :John ;
  owl:assertionProperty :hasParent ;
  owl:targetIndividual :Mary ] .
\end{lstlisting}

From these statements it follows that \texttt{:John :hasParent :Mary} is "false", which makes it a negation of one triple. However, this negation does not have the properties of a full classical negation. The construction only works for a single triple (and not arbitrary collections of triples), and the double negation properties of classical negation are unavailable (which would lead to full FOL expressivity). As for N3Logic, a "false" is available but in a limited form. These types "false" statements can be used as constraints in an ontology, but do not have the properties of a proper classical negation. There is even a more subtle difference between this type of negation and classical negation. The John and Mary example does not state that it is impossible that John has a parent Mary, only that it is not modelled in a particular ontology.  

Universal quantification is available in OWL 2 DL, but in a restricted form. The \texttt{ObjectAllValuesFrom} and \texttt{DataAllValuesFrom} predicates require a set of all individuals to choose from in an quantification. This restricted quantification differs from classical universal quantification, which permits an arbitrary (unlimited) number of individuals.

The Rule Interchange Format (RIF)~\cite{RIF} was an activity started in 2009 within the W3C to develop Web standards for the interchange of rules among disparate systems, especially on the Semantic Web. RIF is a collection of extensible languages and dialects serialized as XML documents. Two types of languages are available: logic-based dialects and dialects for rules with actions. The logic-based dialects include languages based on FOL and various non-FOL semantics, but do not allow negation in the rule head or body (the Horn subset). As far as we know, no RDF syntax was provided for RIF.

Semantic Web Rule Language (SWRL)~\cite{SWRL} is a W3C membership submission that extends the OWL Web Ontology Language with the Rule Interchange Format (RIF). The proposed language extends OWL axioms to include Horn clauses for OWL descriptions and properties and a limited set of built-in functions. This form supports neither the disjunction nor the classical negation of clauses. Also, to guarantee decidability, rules are restricted to only include named individuals (and not existentially introduced individuals). The language allows for explicit quantification, but introducing variables goes beyond RDF semantics~\cite{mei_interpreting_2006}.

The Semantic Web Services Language (SWSL)~\cite{SWSL} is a language for specifying the formal characterizations of  Web service concepts and descriptions of individual Web services. The language consists of two sublanguages: SWSL-FOL, a full FOL language, and SWSL-Rules, a rule-based sublanguage with non-monotonic semantics.  However, the authors of SWSL did not envision the need for full FOL reasoners based on SWSL-FOL as its main use case is creating Web service ontologies. The syntax of SWL is inspired by F-Logic and is not based on RDF.

Several papers by De Bruijn, Tsarkov and Tammet demonstrate the embedding of RDF in FOL using frameworks, such as F-Logic\cite{erdf}, Vampire\cite{tsarkov_2004} and JSON-LD\cite{tammet_2021}. The advantage of FOL lies in its well-established nature and the ability to define its mappings, as demonstrated by these papers. However, our project aims to achieve the expressivity of FOL within the Web language itself, specifically in the RDF model, from the ground up. The papers start from the opposite direction and try incorporating RDF semantics into a framework with FOL semantics. For similar reasons, plain FOL, such as the TPTP language\footnote{\url{https://tptp.org/Proposals/TPILanguage.html}}, is excluded. 

Datalog knows some extensions which aim to incorporate negation into the rule language, most notably semi-positive Datalog and stratified Datalog~\cite{ketsman}. Including negative atoms in rule bodies allows negation to be contained. This inclusion allows for introducing disjunctive and negative statements through a back door. Nonetheless, as is the case for SWRL considered above, Datalog's variants do not allow for existential variables, only to reason over existing ones~\cite{abiteboul}. Furthermore, many Datalog variants are restricted to \textit{safe rules}, where all variables of a rule must occur in a positive atom of the rule body, which further restricts the free use of negative statements.

Answer Set Programming (ASP)~\cite{asp} provides classical "strong" negation and NAF, and even first-order extensions can be written~\cite{lefevre_first_2009}. Eiter et al.~\cite{eiter_combining_2008} provides an extension of ASP with description logic for the Semantic Web. However, using Datalog, ASP, and even Prolog as Web logic creates a dichotomy between the world of data (RDF) and the world of logic (the computer program). In the rationale of RDF Surfaces, it should be possible to negate information in RDF and transport logic/reasoning in a portable way (using RDF). RDF Surfaces could be the language, the syntactic sugar, to transport RDF data with FOL expressivity to an ASP, Datalog, or Prolog program. We do not deny the expressivity of any of these programming languages. Our argument reverses the conventional perspective: Web logics that follow the requirements of the computing agent (decidable, tractable). RDF Surfaces advocates for Web logic for the human agent, emphasizing sharing information and logic in a portable way.

SPARQL can use classical negation and first-order expressivity in filters that can be used to query an RDF data set (that has not FOL expressivity). Three types of negation are supported: (a) the Boolean NOT operator that can be used in filters; (b) the negation as failure operators MINUS and NOT-EXISTS; and (c) a combination of OPTIONAL with the BOUND operator~\cite{angles_negation_2016}. Additionally, SPARQL provides an RDF serialization through SPARQL-SPIN~\cite{knublauch_spin_2013}. However, we regard SPARQL primarily as a query language rather than a Web logic language. 


\section{Existential Graphs}\label{s4}

Existential graphs can be considered a "whiteboard" language for logical reasoning. The whiteboard surface is a logic area that contains thoughts or ideas asserted to be "true". Thoughts or ideas are written on the whiteboard in the form of symbols. These symbols can represent propositions or relations depending on the used diagram system. Groups of symbols can be encircled to create a `nested surface' with special logical properties. Symbols and nested surfaces can be inserted and erased from the whiteboard according to a fixed set of diagram rules. These diagram rules are Peirce symbolic method of natural deduction and represent the calculus of his symbolic language. Symbol manipulation was, for Peirce, the means to make logical reasoning more natural and visual. The dynamic diagrams represent "a moving picture of the actions of the mind in thought"~\cite{sowa_2011}. Peirce developed three diagrammatic systems: the Alpha system, where the symbols represent propositions and the calculus propositional reasoning; the Beta system, where the symbols represent relations and the calculus predicate logic; and the Gamma system, which explores modal and higher-order logic~\cite{sowa_2018}. RDF Surfaces, the core topic of this paper, is the application of existential graphs for RDF. To guide the reader in this translation, a short introduction to the Alpha system and some highlights of the Beta system will be presented below. The application of Peirce's system to RDF will be the topic of the next section.

\subsection{Default positive surface}\label{s4.1}

The \textit{default positive surface} (the whiteboard) is an area that contains zero or more symbols or deeper nested surfaces. In the Alpha system, each symbol represents a proposition. The default surface, or in Peirce's terminology the "sheet of assertion", has the property that any symbol written on it represents a formula that is considered logically "true" (that is, "true" in the relative sense, not an absolute truth). The order in which symbols are written or their position on the surface has no special meaning. The positive surface is interpreted as the logical conjunction ($\land$) of all symbols and surfaces written on it.  If the symbols $A$ and ${B}$ are written on the positive surface, then the conjunction $A \land B$ is interpreted as true by that surface. The empty positive surface is interpreted as a tautology ("true" in every possible interpretation)

\subsection{Negative surface}\label{s4.2}

A \textit{negative surface} (in Peirce's terminology, the "cut") has the property that any symbol that is written on it represents a formula that is considered logically "false". If the symbols $A$ and $B$ are written on a negative surface, then the conjunction $A \land B$ is interpreted as being "false" by that surface. Or stated differently, the negation $\lnot (A \land B)$ is "true" on the default positive surface on which the negative surface is written. A negative surface is the classical negation ($\lnot$) of the symbols written on it. An empty negative surface is interpreted as a contradiction ("false" in every possible interpretation).

\subsection{Nested surfaces}\label{s4.3}

Negative surfaces can be nested inside other negative surfaces, but no nested surfaces are allowed to overlap another. We will define the \textit{surface nesting level} as the number of "negative borders" one must cross to reach a surface's interior. The \textit{parity} of the surface nesting level is the number of crossings modulo 2. We will see that the surface nesting level and its parity will be vital for the diagram manipulation rules, which we will explain in the next subsection. For any surface $S$, the \textit{containment} is defined as the set of all symbols/nested surfaces enclosed by $S$, including the surface $S$ itself.

\begin{figure}[h]
\centering
\includegraphics[width = 250pt]{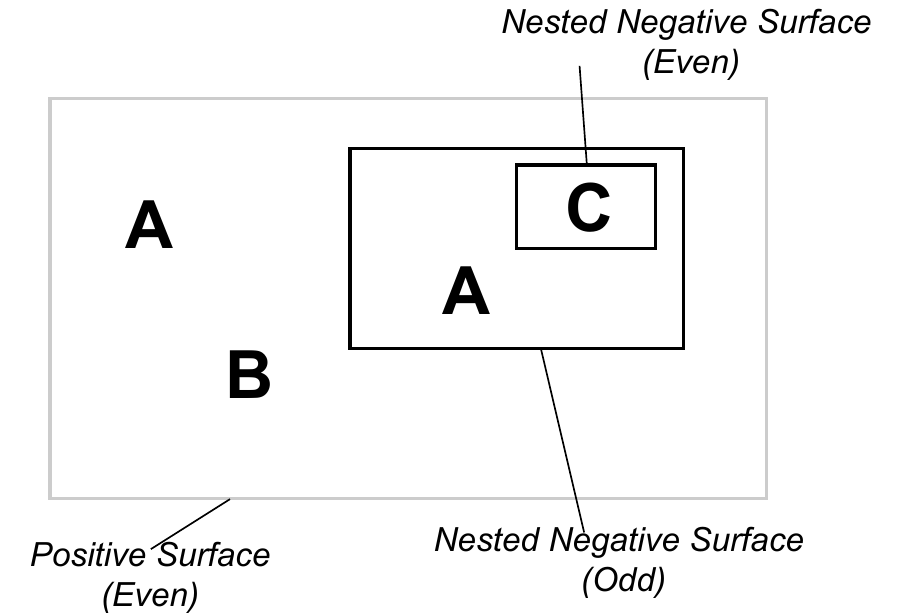}
\caption{A positive surface with the symbols $A$ and $B$, a nested negative surface with the symbol $A$, and a deeper nested negative surface with the symbol $C$. The logical interpretation of this diagram is $A \land B \land \lnot (A \land \lnot C)$. The parity of the surface is the number of "negative borders" one needs to cross to reach the symbols modulo 2. The positive surfaces has parity 0, the negative surfaces with $A$ parity 1, and the negative surface with $C$ parity 0.}
\label{fig:s4_positive_negative}
\end{figure}

\begin{figure}[h]
\centering
\includegraphics[width = 250pt]{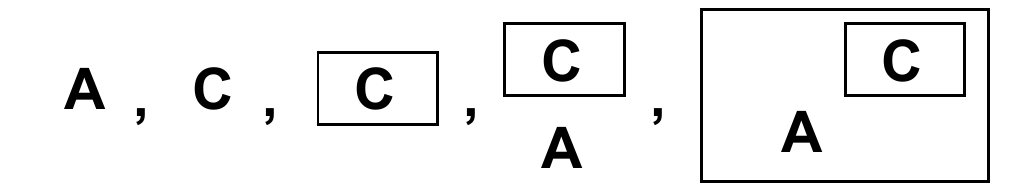}
\caption{The first inner nested negative surface (the one with odd parity) of \cref{fig:s4_positive_negative} has a containment of 5 symbols/nested surfaces.}
\label{fig:s4_containment}
\end{figure}

Positive surfaces, negative surfaces, and nesting are all required to interpret any composite diagram. As an example, in \cref{fig:s4_positive_negative}, we see a default positive surface with propositions $A$ and $B$, a nested negative surface with proposition $A$, and inside the nested surface another nested surface with proposition $C$. On both the default positive surface and the nested negative surface, $A$ represents the same proposition. The positive surface has parity 0, the first nested negative surface has parity 1, and the innermost negative surface has parity 0. The full interpretation of this diagram is $A \land B \land \lnot ( A \land \lnot C)$. The containment of the first (outer) negative surface in~\cref{fig:s4_positive_negative} includes 5 symbols/nested surfaces as is shown in \cref{fig:s4_containment}.
 
\subsection{Diagram rules}\label{s4.4}

Peirce provided for his Alpha system four diagram manipulation rules to insert and erase symbols and nested surfaces:

\begin{itemize}
    \item \textbf{R1 Insertion}: Any symbol/nested surface can be introduced on any surface with parity 1.
    \item \textbf{R2 Erasure}: Any symbol/nested surface can be erased on any surface with parity 0.
    \item \textbf{R3 Double Cut}: A double nested surface can be replaced by its interior when the outer region is empty.
    \item \textbf{R4 (De)iterate}: Any symbol/nested surface $S'$ on a surface $S$ can be placed or erased from any surface that is not part of $S'$, but contained by $S$.
\end{itemize}

\begin{figure}[h]
\centering
\includegraphics[width = 200pt]{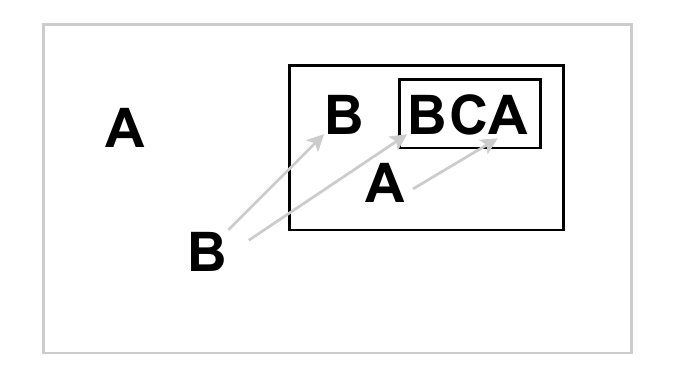}
\caption{Using rule \textbf{R4} a copy of a symbol or nested surface can be placed in at any surface level which is contained by the origin surface. But, it is not possible to place a copy of a nested surface within itself.}
\label{fig:s44_deiteration_example}
\end{figure}

The last rule \textbf{R4} requires some explanation. A copy of any symbol/nested graph can be added or erased from any nesting level. Starting from \cref{fig:s4_positive_negative} we can create \cref{fig:s44_deiteration_example} by adding the $B$ symbol in the nested surface with party 1, and the inner nested surface with party 0. Likewise, the $A$ symbols can be added at any nesting level. Rule \textbf{R4} also erases these $A$ and $B$ copies. Rule \textbf{R4} does not allow to place a copy of a nested surface within itself.

\begin{figure}[h]
\centering
\includegraphics[width = 250pt]{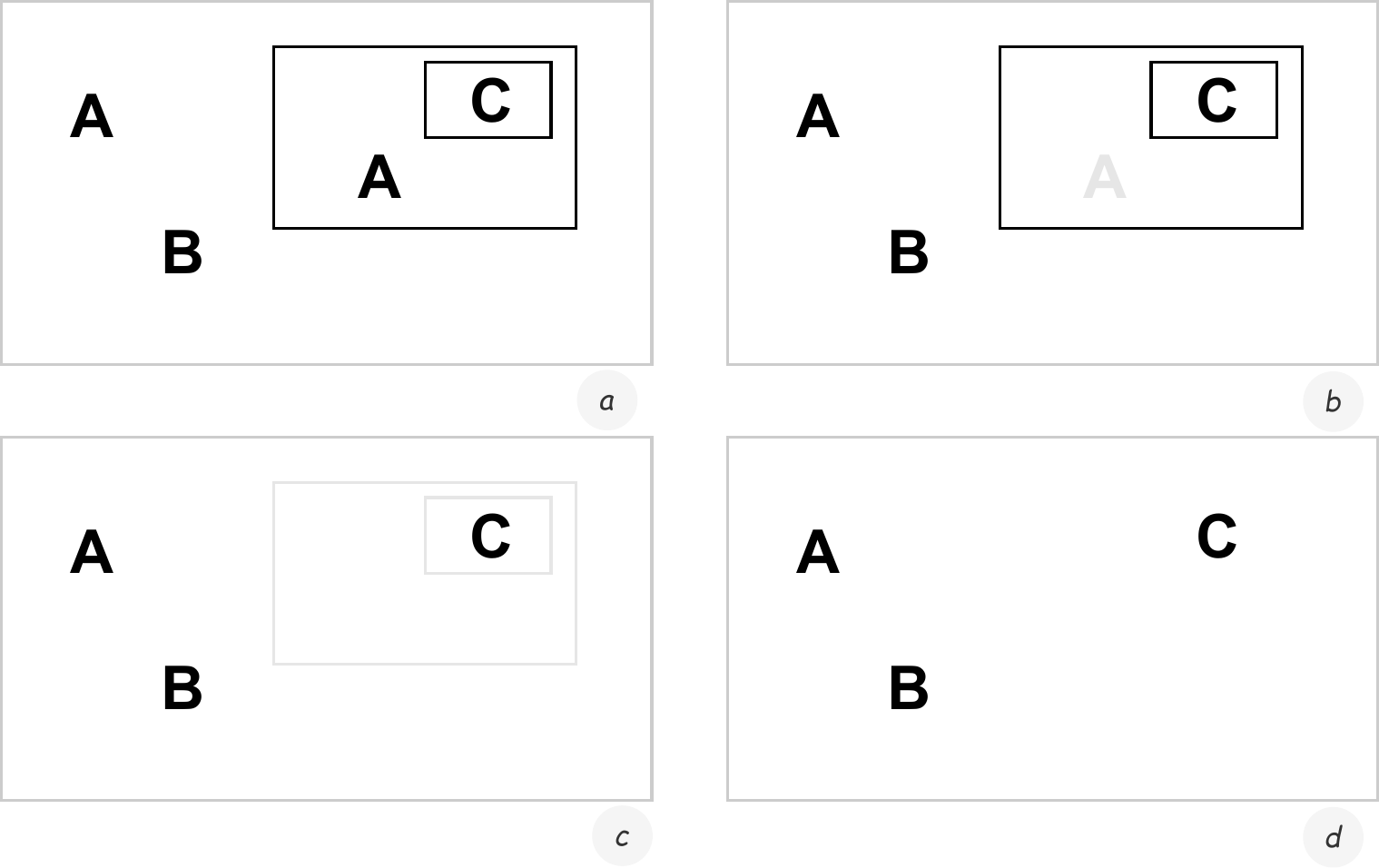}
\caption{(a) A representation of $A \land B \land \lnot ( A \land \lnot C )$ ; (b) the deiterate rule \textbf{R4} is applied to erase a copy of $A$ from the nested negative surface; (c) the double cut rule \textbf{R3} is applied to erase the double nested surface; (d) the result of the deduction $A \land B \land C$.}
\label{fig:s44_deiteration}
\end{figure}

Diagram manipulation rules can be applied to create new diagrams that maintain the same logical truth value as the original diagram. For instance, \cref{fig:s44_deiteration}a is a copy of \cref{fig:s4_positive_negative}. By applying the deiterate rule \textbf{R4} to this diagram, we can remove the $A$ copy from the nested negative surface, which results in \cref{fig:s44_deiteration}b. As a next step, we see that \cref{fig:s44_deiteration}b contains a double nested negative surface, which can be erased using the double cut rule \textbf{R3}. This results in \cref{fig:s44_deiteration}c and eventually  \cref{fig:s44_deiteration}d. This symbolic calculus provides a deduction for \cref{eq:s44_deduction}.

\begin{equation}\label{eq:s44_deduction}
A \land B \land \lnot ( A \land \lnot C ) \overset{R4}{\vdash}  A \land B \land \lnot(\lnot C)  \overset{R3}{\vdash} A \land B \land C
\end{equation}

An implication can be recognized in \cref{eq:s44_deduction}. A negative surface containing $A$ plus a deeper nested negative surface containing $C$ is the diagrammatic equivalent of the logical formula $\lnot (A \land \lnot C)$, which is, by definition, the implication: $\lnot (A \land \lnot C) \equiv A \rightarrow C$. That is, it cannot be the case that $A$ is "true" and $C$ is not "true". Stated differently, from $A$ follows $C$. This can be written in symbolic form as:

\begin{equation}
A \land B \land \lnot ( A \land \lnot C ) \vdash 
A \land B \land ( A \rightarrow C ) \vdash
A \land B \land C
\end{equation}

\begin{figure}[h]
\centering
\includegraphics[width = 250pt]{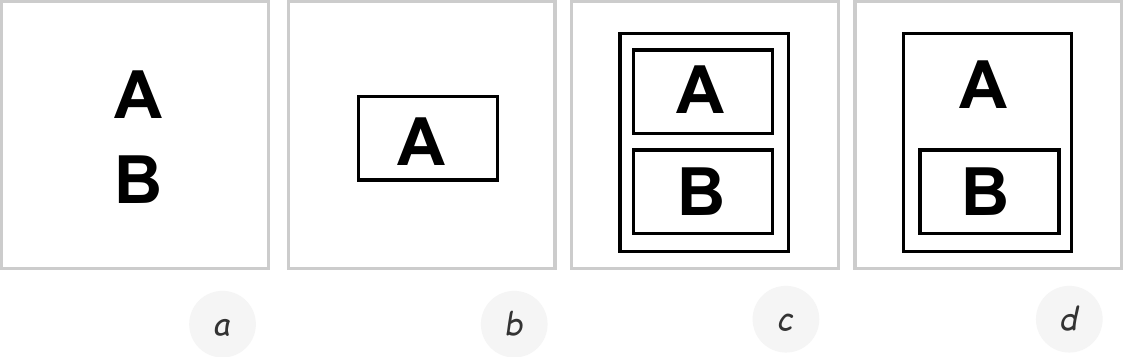}
\caption{Diagram representations of (a) $A \land B$, (b) $\lnot A$, (c) $A \lor B$, and (d) $A \rightarrow B$}
\label{fig:s44_compound}
\end{figure}

The correspondence between Alpha calculus and propositional logic (Beta calculus with predicate logic) is not coincidental. Zeman formally established it in the early 1960s~\cite{zeman_graphical_1964}. Important for our discussion is that any compound truth-functional statement can be written diagrammatically using symbols and (nested) negative surfaces:

\begin{itemize}
 \item The logical conjunction $\land$ is given by writing symbols (or nested negative surfaces) on a positive surface (\cref{fig:s44_compound}a).
 \item The logical negation $\lnot$ is given by writing symbols (or nested negative surfaces) on a negative surface  (\cref{fig:s44_compound}b).
 \item The logical disjunction $\lor$ is given by noting that $A \lor B \equiv \lnot ( \lnot A \land \lnot B)$  (\cref{fig:s44_compound}c).
 \item The logical implication $\rightarrow$ is given by noting that $A \rightarrow B \equiv \lnot ( A \land \lnot B)$  (\cref{fig:s44_compound}d).
\end{itemize}

\subsection{Quantification}\label{s4.5}

\begin{figure}[h]
\centering
\includegraphics[width = 250pt]{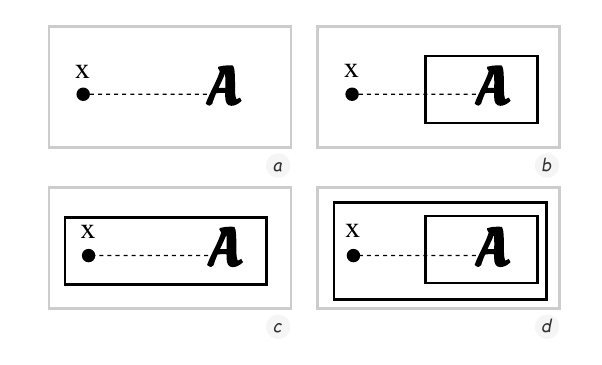}
\caption{Quantification using the Beta system. The interpretations of these diagrams are: (a) $\exists x: \mathcal{A}(x)$, (b) $\exists x: \lnot\mathcal{A}(x)$, (c) $\lnot (\exists x: \mathcal{A}) \equiv \forall x : \lnot\mathcal{A}(x)$, and (d) $\lnot (\exists x: \lnot\mathcal{A}) \equiv \forall x : \mathcal{A}(x)$.}
\label{fig:s45_quantification}
\end{figure}

The diagrams thus far addressed reasoning with propositions using the Alpha system. The Beta system expands the Alpha system by adding quantification and the logic of relations. In this article, we will only provide a concise example of the Beta system, providing enough detail to understand the quantification techniques of the next section. 

A (possibly indexed) heavy dot $\bullet$ denotes that some identity exists in the Beta system. These heavy dots can be extended into a "line of identity", which has the same interpretation as the heavy dot. When two or more heavy dots appear in a diagram with indexes with equal labels, they act as coreferences and denote the same identity. In the Beta system, symbols are interpreted as relations. A relation's \textit{arity} is the number of "lines of identity" attached to the symbol. In our examples, we will use a script-like font to differentiate relation symbols from proposition symbols.

In \cref{fig:s45_quantification}, we see a Beta diagram that means "There exists some entity x which is an $\mathcal{A}$" or as a logical formula $\exists x : \mathcal{A}(x)$. If the relation symbol is written on a negative surface, as in \cref{fig:s45_quantification}b, the interpretation becomes "There exists some entity x which is not an $\mathcal{A}$" or as a logical formula $\exists x : \lnot \mathcal{A}(x)$. When the heavy dot is placed inside the nested surface, as in \cref{fig:s45_quantification}c, the contents of the surface and the heavy dot will be denied. This diagram means "It is false that some entity x is an $\mathcal{A}$," which is equivalent to "Nothing is an $\mathcal{A}$," or as a logical formula $\forall x : \lnot \mathcal{A}(x)$. When we move the relation to a deeper nested negative surface, as in \cref{fig:s45_quantification}d, we get the interpretation "It is false that some identity $x$ is not an $\mathcal{A}$," or stated differently "Every entity $x$ is an $\mathcal{A}$", which can be written as the logical formula $\forall x : \mathcal{A}(x)$.  

De Morgan's identity, \cref{eq:s45_demorgan}, can be recognized in these four examples. A heavy dot on a surface with parity 0 (e.g., the default positive surface) should be interpreted as an existentially quantified variable. A heavy dot on a surface with parity 1 should be interpreted as a negated existentially quantified variable equivalent to a universally quantified variable.

\begin{equation}\label{eq:s45_demorgan}
   \lnot(\exists x : \mathcal{P}(x)) \equiv \forall x : \lnot \mathcal{P}(x)
\end{equation}

\begin{figure}[h]
\centering
\includegraphics[width = 150pt]{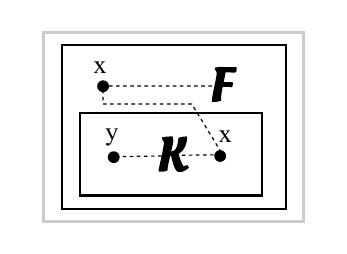}
\caption{The interpretation of the quantified variables $x$ and $y$ depend on the surface nesting level. The outermost $x$ is on the negative surface with nesting level 1, thus, its scope is the outer surface. The $y$ is on the negative surface with nesting level 2, and its scope is the inner surface.}
\label{fig:s45_quantification_2}
\end{figure}

Using the Beta system, arbitrary complex statements with predicates can be composed of (indexed) heavy dots (extended into "lines of identity") and relation symbols. \cref{fig:s45_quantification_2} expresses the implication "It is false that some entity x is famous ($\mathcal{F}$) and not some entity y knows ($\mathcal{K}$) x". That is, "For any entity x that is famous, there is some entity y that knows x." Or, stated as a logical formula: 

\begin{equation}
\lnot ( \exists x :  \mathcal{F}(x) \land \lnot ( \exists y: \mathcal{K}(y,x) ) ) \\
\equiv \\
\forall x \exists y : \mathcal{F}(x) \rightarrow \mathcal{K}(y,x)
\end{equation}

The scope of the heavy dot depends on the surface with the lowest surface nesting level on which the (possibly indexed) heavy dot is written. In \cref{fig:s45_quantification_2}, the heavy dot with index $x$ is placed on the negative surface with level 1 and extended to the negative surface with level 2. 

For the Beta system diagram, rules can be provided that are an extension of the four Alpha system diagram rules that were described in \cref{s4.4}, by including rules for heavy dots. For the remainder of this paper, we do not delve into the specifics of this extension. Instead, we refer the reader to Roberts~\cite{roberts_existential_1973} for a thorough introduction. It is sufficient to state that a heavy dot is treated as a wildcard. 

\begin{figure}[h]
\centering
\includegraphics[width = 350pt]{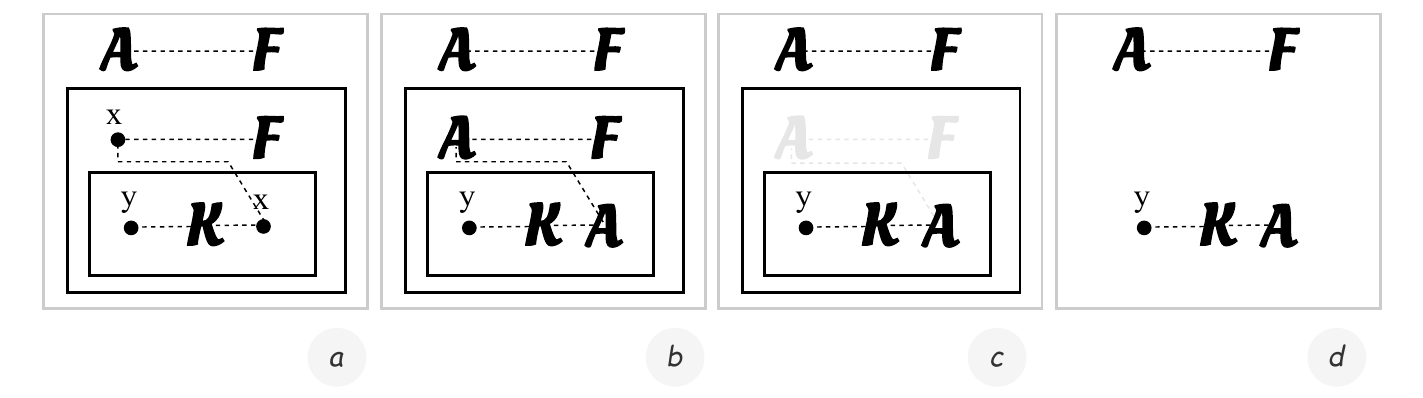}
\caption{Concise example of a derivation using the Beta system: (a) the original diagram, (b) $x$ is a wildcard which matches $A$ on the positive surface, (c) using the deiterate rule \textbf{R4} the copy $\mathcal{A}...\mathcal{F}$ can be removed from the nested negative surface, (d) using the double cut rule \textbf{R3} the double nested surface can be removed.}
\label{fig:s45_quantification_3}
\end{figure}

In \cref{fig:s45_quantification_3}, we see an example of a Beta system derivation using heavy dots. \cref{fig:s45_quantification_3}a is a copy of \cref{fig:s45_quantification_2}, which includes on the default positive surface the statement $\mathcal{A}...\mathcal{F}$ that symbolizes: "There is some A which is famous". In \cref{fig:s45_quantification_3}b, we see that a heavy dot was applied as a wildcard: $x...\mathcal{F} = \mathcal{A}...\mathcal{F}$ if $x = \mathcal{A}$.  This replacement  $x = \mathcal{A}$ must also be done for the $x$ in the deeper nested negative surface.  In \cref{fig:s45_quantification_3}c, we apply the deiteration rule \textbf{R4} to remove the copy $\mathcal{A}...\mathcal{F}$ from the nested negative surface. From \cref{fig:s45_quantification_3}c, we get to \cref{fig:s45_quantification_3}d by applying the double cut rule \textbf{R3}. 

It should be noted that the Beta system does not have a notion of logical constants. The diagram notation $\mathcal{A}...\mathcal{F}$ should be interpreted as:

\begin{equation}
\exists x : \mathcal{A}(x) \land \mathcal{F}(x)
\end{equation}

In the Beta system, relations with any arity can be introduced on the whiteboard (the $\mathcal{A}$ and $\mathcal{F}$ in our examples were of arity 1 and $\mathcal{K}$ of arity 2).  We will see in the next section on RDF Surfaces we use Peirce's Existential graphs as an inspiration for applying surface language in RDF context, but we will deviate from the Beta system and allow constants, use only relations of arity 2, and reinterpret the heavy dot as the blank nodes of RDF.


\section{RDF Surfaces}\label{s5}

In the previous section provided the foundation for expressing FOL using Peircian diagram logic. This section applies Peirce diagrams to the Semantic Web. As an introductory step, we add RDF triples in the form of an implication in a Peirce diagram and demonstrate that the same diagram rules of \cref{s4} can be applied to derive new RDF triples. In a second step, we translate the diagrams into an RDF syntax and semantics we named \textit{RDF Surfaces}.

With existential graphs, we are not limited to including only abstract symbols in the diagrams. In \cref{fig:s4_positive_negative}, the $A$, $B$, and $C$ can be replaced by any concrete object, such as ideas, drawings, and even RDF statements and still retain the same expressive power of visualized reasoning. Using these insights, we use the results of our previous section to apply existential graphs to RDF following the BLOGIC vision of Hayes. This new form, called RDF Surfaces, will demonstrate how the expressivity of FOL can be implemented with only two additions to the RDF Model: \textit{surfaces} on which to draw RDF Graphs and the explicit scoping of quantified variables using collections of blank nodes, called the \textit{graffiti}, which are the "heavy dots" of the previous section.

\begin{figure}[h]
\centering
\includegraphics[width = 250pt]{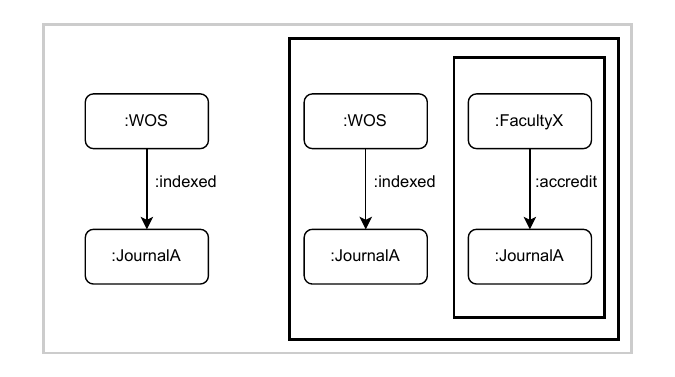}
\caption{A diagram representations of the RDF fact \texttt{$\langle$:WOS :indexed :JournalA$\rangle$} and the  implication \texttt{$\langle$:WOS :indexed :JournalA$\rangle$ $\rightarrow$ $\langle$:FacultyX :accredit :JournalA$\rangle$}. }
\label{fig:s5_positive_negative_rdf}
\end{figure}

To illustrate our approach, \cref{fig:s5_positive_negative_rdf} presents an example where the previous section's abstract symbols $A$, $B$, and $C$ have been replaced with RDF statements. The default surface in our example contains the diagram:

\begin{equation*}
     \Ovalbox{\texttt{:WOS}} \stackon{\textrightarrow}{\small{\texttt{:indexed}}} \Ovalbox{\texttt{:JournalA}}      
\end{equation*}
 
This diagram stands for the RDF triple \texttt{$\langle$:WOS :indexed :JournalA$\rangle$}, i.e.~"WOS added JournalA to its index". It also contains an implication with the meaning "If WOS indexed JournalA, then FacultyX accredited JournalA". As such, \cref{fig:s5_positive_negative_rdf}  expressed in symbolic form represents:

\begin{equation}
   \begin{split}
    & \texttt{$\langle$:WOS :indexed :JournalA$\rangle$ $\land$ } \\
    & \big( \texttt{$\langle$:WOS :indexed :JournalA$\rangle$ $\rightarrow$ $\langle$:FacultyX :accredit :JournalA$\rangle$} \big)
   \end{split} 
\end{equation}

Using the diagram rules from \cref{s4.4} we can deduce \texttt{$\langle$:FacultyX :accredit :JournalA$\rangle$} in two steps:

\begin{itemize}
    \item \textbf{R4} : deiterate \Ovalbox{\texttt{:WOS}} \stackon{\textrightarrow}{\small{\texttt{:indexed}}} \Ovalbox{\texttt{:JournalA}} from the nested negative surface with nesting level 1.
    \item \textbf{R3}: remove the double cut that was created by the first step.
\end{itemize}

To make RDF Surfaces more concrete, we define the concept of a \textit{Hayes triple} and a \textit{Hayes graph}.

\begin{definition}
 \label{s5.def.1}
A \textit{Hayes triple} consists of three components $\langle \pmb{Gr}$ $\pmb{S}$ $\pmb{H} \rangle$ where:
 \begin{itemize}
     \item $\pmb{Gr}$ is a (possibly empty) set of blank nodes, called \textit{graffiti}.
     \item $\pmb{S}$ is a surface type.
     \item $\pmb{H}$ is a (possibly empty) set of RDF triples and Hayes triples, called the \textit{Hayes graph}.
 \end{itemize}
\end{definition}

Following Viswanathan~\cite{viswanathan_cs498mv_2018}, we can define a scope, plus bound and free occurrences of graffiti nodes, as follows:

\begin{definition}
 \label{s5.def.2}
For the \textit{Hayes triple} $\langle\pmb{Gr}$ $\pmb{S}$ $\pmb{H}\rangle$,  $\pmb{H}$ is said to be the \textit{scope} of the graffiti nodes in $\pmb{Gr}$.
\end{definition}

\begin{definition}
 \label{s5.def.3}
Every occurrence of a graffiti node $g$ in $\pmb{Gr}$ as a blank node in any (nested) RDF triple of $\pmb{H}$ is called a \textit{bound} occurrence of $g$ in $\pmb{H}$. The blank node in that RDF triple is a \textit{coreference} to $g$. Any occurrence of a graffiti $g$ that is not bound is called a \textit{free} occurrence of $g$ in $\pmb{H}$.
\end{definition}

We say that the graffiti nodes of $\pmb{Gr}$ are \textit{on} the surface defined by the Hayes graph $\pmb{H}$. The recursive definition of a Hayes graph reflects the fact that, following \cref{s4}, we deal with nested surfaces. The surface type translates to different surface interpretations we saw in \cref{s4}, i.e.~positive and negative. In general, more surface types can be imagined. Hayes BLOGIC presentation, for example, mentioned neutral and deprecated surface types as possible extensions. Neutral surfaces would not be asserted or negated and could be used for packaging RDF triples without giving them a truth value. Deprecated surfaces could be used to provide a time constraint on the truth value of a surface. 

Graffiti, in the form of blank nodes, are a direct translation of the heavy dots of the previous section. Blank nodes that appear in the RDF triples of $\pmb{H}$ act as coreferences to the graffiti nodes $\pmb{Gr}$ with the same label on an ancestor surface. These graffiti nodes are scoped in $\pmb{H}$ in such a way that if a deeper nested Hayes graph exists that contains graffiti nodes using the same label as a parent Hayes graph, this deeper nested Hayes graph creates a new logical scope for the graffiti node with that label. If a blank node in a RDF triple does not have such a coreference to a graffiti node, it is said to be free.

\begin{figure}[h]
\centering
\includegraphics[width = 250pt]{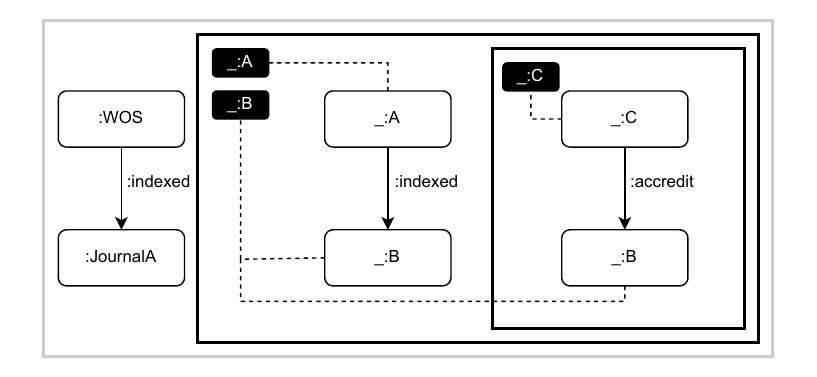}
\caption{A diagram representation of the fact \texttt{:WOS :indexed :JournalA} and an implication which means "Every \texttt{A} that indexed \texttt{B}, has some \texttt{C} that accredits \texttt{B}".}
\label{fig:s5_graffiti_rdf}
\end{figure}

\cref{fig:s5_graffiti_rdf} provides an example. A graffiti node with label \texttt{\_:B} is written on the negative surface with parity 1. There are also blank nodes written on the same surface, and on a deeper nested negative surface with parity 0 with the same label \texttt{\_:B}. All these blank nodes, labeled \texttt{\_:B}, are coreferences to the graffiti node with label \texttt{\_:B}.

Given the concept of a Hayes graph we can define an RDF Surface.

\begin{definition}
 \label{s5.def.4}
An \textit{RDF Surface} is a Hayes triple with a \textit{positive} surface type, and every Hayes triple nested within the RDF Surface has a \textit{negative} surface type (regardless of the depth of their nesting).
\end{definition}

This definition reflects the role of the default surface as a positive surface, i.e.~its content has the value "true". All the nested surfaces are negative, i.e.~its content has the value "false". 

Each blank node that occurs somewhere in an RDF Surfaces graph, i.e.~either in the Hayes graph of the positive surface or in that of a nested Hayes graph, but not in any of the sets of graffiti of its parent Hayes graphs, is considered \textit{free}. We subsequently 'existentially close' the RDF Surfaces graph by adding graffiti nodes on the top level, the default positive surface, for every freely occurring blank node. In the case of identically named blank nodes, there is only one single graffiti to which all occurrences are referenced. This coincides with the approach taken in plain RDF, i.e.~any blank node in RDF is implicitly assumed to be existentially quantified. Within RDF Surfaces, we aim to support any existing RDF document. To achieve this, we create an implicit existential closure by capturing these seemingly 'free' blank nodes within RDF in the set of graffiti of the top-level, positive Hayes graph, 'pinning down' their existential nature.  

An example of a full RDF Surfaces graph using explicit quantification can be seen in \cref{fig:s5_graffiti_rdf}. The default surface contains the assertion: \texttt{$\langle$:WOS :indexed :JournalA$\rangle$}. The right part of the figure contains an implication similar to the one in  \cref{fig:s5_positive_negative_rdf}, but now using the graffiti blank nodes (in black) to denote quantifications that are used in RDF Surfaces. The graffiti nodes \texttt{\_:A} and \texttt{\_:B} are attached to the negative surface with nesting level 1,  parity 1. This position of the graffiti nodes \texttt{\_:A} and \texttt{\_:B} on a negative surface with parity 1 provides an interpretation of these nodes as \textit{existential} quantified variables that can be transformed into \textit{universal} quantified variables. However, the graffiti node \texttt{\_:C} is attached to a negative surface with nesting level 2, parity 0. The position of the graffiti node \texttt{\_:C} on this negative surface with parity 0 interprets the node as an \textit{existential} quantified variable. All other blank nodes (in white) on the negative surface and its deeper levels are coreferences to the graffiti nodes with the same label. The meaning of implication in \cref{fig:s5_graffiti_rdf} is  "Every \texttt{A} that indexed \texttt{B}, has some \texttt{C} that accredits \texttt{B}". The full diagram can be symbolized using \cref{eq:5_to_path}, as follows:

\begin{equation}
\label{eq:5_to_path}
 \begin{split}
   &\texttt{$\langle$:WOS :indexed :JournalA$\rangle$} \ \land  \\
   & ( \forall a,b : \  \exists c : \texttt{$\langle a$ :indexed $b \rangle$} \rightarrow \texttt{$\langle c$ :accredit $b \rangle$} )
 \end{split}
\end{equation}

Applying the diagram rules of \cref{s4.4} in combination with the extension for quantification in \cref{s4.5}, the following deduction is possible:

\begin{itemize}
 \item \Ovalbox{\texttt{\_:A}} \stackon{\textrightarrow}{\small{\texttt{:indexed}}} \Ovalbox{\texttt{\_:B}} can be deiterated using rule \textbf{R4} when \texttt{\_:A} is equal to \texttt{:WOS}, and \texttt{\_:B} is equal to \texttt{:JournalA}.	
 \item This leaves us with a double cut that can be removed using rule \textbf{R3}.
 \item The result is a graffiti node labeled \texttt{\_:C} on the default positive surface and the new fact:\\  \Ovalbox{\texttt{\_:C}} \stackon{\textrightarrow}{\small{\texttt{:accredit}}} \Ovalbox{\texttt{:JournalA}} .
\end{itemize}

When multiple facts are available, the diagram rules from \cref{s4.4} can be repeatedly applied to make derivations, which always result in the same graffiti nodes labeled $\texttt{\_:C}$. Using the iteration rule \textbf{R4} we can add many copies of the implication of \cref{fig:s5_graffiti_rdf}. 

\subsection{Notation3 serialization}\label{s5.1}

Hayes introduced an annotated Turtle syntax, using comments, in his BLOGIC presentation to express Peirce's diagrams. With RDF Surfaces we opted to use a subset of the Notation3 (N3)~\cite{n3spec} syntax to provide a more structured serialization format in RDF itself. N3 already has parsers, such as EYE~\cite{eye} and n3~\cite{verborgh_n3_2024}, that provide syntactical support for graph terms and RDF lists as first-class citizens of the language, which provide a possibility for a direct translation of RDF Surfaces diagrams into N3. Graph terms can be used to describe a RDF/Hayes graph $\pmb{H}$ on a surface and lists for the set $\pmb{Gr}$ of graffiti on a surface. 

It is important to differentiate between the requirements for the RDF model (and its semantics) to describe RDF Surfaces and the serialization format to transport that model. 

Our first choice for using N3 as a serialization format is pragmatic, based on de facto approaches to make statements about sets of RDF triples. Alternatively, we could have considered named graphs. However, if the surfaces have IRIs, this approach risks self-referential surfaces and paradoxes. If the surfaces use blank node identifiers, it also leads to issues with the quantification of surfaces. We aimed to stay close to the Semantic Web ideal that everything can be expressed as triples.

Our requirements for the RDF model focus on identifying the minimal additions needed to achieve the expressivity of FOL with explicit quantification. RDF Surfaces requires two additions: a concept of a surface containing a (possible empty) set of triples, and a concept of a collection of graffiti blank nodes that act as existentially quantified variables. RDF Surfaces relies on the simplest version of entailment: only Simple Entailment is required\footnote{\url{https://www.w3.org/TR/rdf11-mt/\#simpleentailment}} and other entailments such as RDFS follow by applying RDF Surfaces formulas. That is, from:

\begin{equation*}
\texttt{$\langle$:CelestialObject rdfs:subClassOf :Cheese$\rangle$} \land
 \texttt{$\langle$:Moon a :CelestialObject$\rangle$}
\end{equation*}

one may conclude

\begin{equation*}
 \texttt{$\langle$:Moon a :CelestialObject$\rangle$}
\end{equation*}

but not

\begin{equation*}
 \texttt{$\langle$:Moon a :Cheese$\rangle$}, 
\end{equation*}

as this relies on RDFS entailment. In RDF Surfaces, we build RDF entailment from the ground up, creating a foundational layer to make entailments explicit, based on FOL. To entail the RDFS version, an extra formula can be introduced that explains what \texttt{rdfs:subClassOf} means, such as the RDF Surfaces version of the symbolic form:

\begin{equation*}
\forall x,y,z : (
  \texttt{$\langle$x rdfs:subClassOf y$\rangle$} \land
  \texttt{$\langle$z a x$\rangle$} ) \rightarrow
  \texttt{$\langle$z a y$\rangle$} .
\end{equation*}

To allow for a serialization of surfaces and graffiti we use only a subset of the N3 \textit{syntax} (not its semantics):

\begin{itemize}
 \item The Turtle textual syntax for RDF: meaning that any valid RDF 1.1 Turtle~\cite{Turtle} graph is a valid RDF Surfaces graph.
 \item N3 list terms as a set of graffiti (blank) nodes. 
 \item N3 graph terms as a conjunction of quoted statements. 
 \item A new IRI \texttt{log:onNegativeSurface} which indicates the type of surface (currenly only the negative surface type is used).
\end{itemize}

A (default) positive surface is implicit assumed on the boundary of every RDF document. That is, every RDF triple in existing RDF documents is assumed to be "on" a (default) positive surface. Every 'free' blank node in an RDF graph is assumed to be implicit existential closed in the set of graffiti nodes of the (default) positive surface. These graffiti nodes cannot be shared between RDF documents. When combining two or more RDF documents, a new set of graffiti nodes must be created ("engraved" in Hayes terminology) in the resulting RDF document.

A Hayes triple is expressed in "RDF Surfaces in N3" (N3S) as a Turtle/RDF triple with on the subject position an N3 collection to represent the graffiti nodes that are  "on" a surface. In the predicate position, we write the surface type: in our case \texttt{log:onNegativeSurface}. In the object position, an N3 graph term represents the Hayes graph. The usage of N3 graph terms is restricted so that they only occur in the object position of a triple having a surface type as a predicate. N3 collections (lists terms) are first-class citizens in N3 (and N3S). We require this feature in the N3 serialization of RDF Surfaces because "classical" RDF lists introduced blank nodes. This is problematic, as we need to exactly know where blank nodes are quantified.

\cref{list:graffiti_rdf} provides an example of such a N3S serialization of the RDF Surfaces diagram of \cref{fig:s5_graffiti_rdf}. The triple on line 4 is "on" the (default) positive surface. Line 6 defines a negative surface with the graffiti nodes \texttt{\_:A} and \texttt{\_:B}, interpreted to be \textit{on} the negative surface. The blank nodes \texttt{\_:A} and \texttt{\_:B} on lines 8 and 11 act as coreferences to the graffiti nodes declared by the outer negative surface. Line 10 defines a nested negative surface with one graffiti node, \texttt{\_:C}. The blank node \texttt{\_:C} on line 11 is a coreference to the graffiti node declared on line 10.

\begin{lstlisting}[numbers=left,label=list:graffiti_rdf,caption={The N3S serialization of the RDF Surfaces diagram depicted in \cref{fig:s5_graffiti_rdf}.}]
@prefix log: <http://www.w3.org/2000/10/swap/log#> .
@prefix : <https://example.org/ns#> .

:WOS :indexed :JournalA .

( _:A _:B ) log:onNegativeSurface {

    _:A :indexed _:B . 
     
    ( _:C ) log:onNegativeSurface {
        _:C :accredit _:B .
    } .
} .
\end{lstlisting}

\subsection{Blank nodes and explicit scoping of logical variables}\label{s5.2}

Using the concept of scope, as defined by \cref{s5.def.2}, a careful relabeling of graffiti nodes in \cref{list:graffiti_rdf} without changing the meaning of the RDF Surfaces graph is possible. 

\begin{lstlisting}[numbers=left,label=list:graffiti_rdf_b,caption={A relabeled version of \cref{list:graffiti_rdf} with the same meaning, using the scoping of graffiti nodes.}]
@prefix log: <http://www.w3.org/2000/10/swap/log#> .
@prefix : <https://example.org/ns#> .

:WOS :indexed :JournalA .

( _:A _:B ) log:onNegativeSurface {

    _:A :indexed _:B . 
     
    ( _:A ) log:onNegativeSurface {
        _:A :accredit _:B .
    } .
} .
\end{lstlisting}

In \cref{list:graffiti_rdf_b}, a graffiti node with label \textit{\_:A} appears on lines 6 and 10, with corresponding blank nodes on lines 8 and 11. The question remains: which graffiti nodes on lines 6 and 10 should bind? RDF Surfaces uses the following \textit{binding convention}:

\vspace{10px}
\textbf{Convention}:
 Blank nodes bind to graffiti nodes with the same label on the closest parent surface.
\vspace{10px}

In our example, this would imply that the blank node with label \texttt{\_:A} on line 8 is bound to the graffiti node with the same label declared on line 6, and the blank node with label \texttt{\_:A} on line 11 is bound to the graffiti node with the same label declared on line 10. This convention makes the meaning of \cref{list:graffiti_rdf_b} equal to \cref{list:graffiti_rdf}.

The binding of blank nodes that are \textit{not} mentioned in any graffiti is implicit and becomes `existentially closed' on the top-level (default) positive surface. In this sense, RDF Surfaces mimic the behavior of plain RDF. Well-known ambiguities for quantification can be avoided~\cite{hogan_2014}. For example, the design documents for N3Logic added syntactical features, such as \texttt{@forAll} and \texttt{@forSome} for quantification, but because graphs have no order in RDF, ambiguity exists in which N3 formulas are quantified by these quantifiers~\cite{arndt_implicit_2019}. In RDF Surfaces, quantification is unambiguous using mandatory graffiti nodes with scope.

\subsection{Explicit negation}\label{s5.3}

Within the boundaries of RDF Surfaces, positive or negative facts can be stated without requiring access to all possible RDF triples in a much bigger world. For instance, to express that an RDF triple in an RDF Surfaces graph does not have a particular property, we can state (\cref{list:wos-not-abc}): 

\begin{lstlisting}[numbers=left,label=list:wos-not-abc,caption={An RDF Surfaces graph with an explicit negation with meaning "WOS did not index JournalABC."}]
@prefix log: <http://www.w3.org/2000/10/swap/log#> .
@prefix : <https://example.org/ns#> .

() log:onNegativeSurface {
    :WOS :indexed :JournalABC .
} .
\end{lstlisting}

The meaning of \cref{list:wos-not-abc} is: ``It is not the case that \texttt{:WOS :indexed :JournalABC}''. If this RDF Surfaces graph would contain any triples stating that \texttt{:WOS :indexed :JournalABC}, or if this could be inferred by deduction, it would be in conflict with the stated negation and thus create a contradiction. These contradictions are explicitly available by the semantics of RDF Surfaces and are not hidden as in the current RDF model.

\begin{figure}[h]
\centering
\includegraphics[width = 290pt]{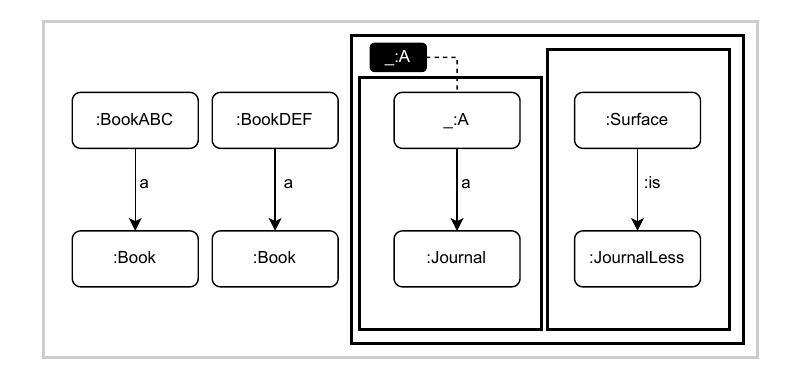}
\caption{A diagram representations of \cref{list:no-naf}. There is no negative fact on the default positive surface to deiterate the negation of \texttt{$\langle$ \_:A a :Journal $\rangle$}. It is \textbf{not} possible to conclude \texttt{$\langle$ :Surface :is :JournalLess $\rangle$}.}
\label{fig:s5_naf}
\end{figure}

Adding explicit negation as data or as a consequence of an implication makes it possible to state which statements are "false". This explicit negation is not equivalent to NAF. One could be tempted to create a negation to simulate NAF as in \cref{list:no-naf}. However, \cref{list:no-naf} does not entail \texttt{:Surface :is :JournalLess}. The RDF Surfaces graph, as shown in \cref{fig:s5_naf}, does not contain the \textit{explicit (negative) fact} that some entity is \textit{not} a journal. The deiteration rule \textbf{R4} cannot be applied.

\begin{lstlisting}[numbers=left,label=list:no-naf,caption={A negative surface \textbf{cannot} be used as a negation as failure. The triple \texttt{$\langle$:Surface :is :JournalLess$\rangle$} is \textbf{not} a logical consequence of this RDF Surfaces graph.}]
@prefix log: <http://www.w3.org/2000/10/swap/log#> .
@prefix : <https://example.org/ns#> .

:BookABC a :Book .
:BookDEF a :Book .

(_:A) log:onNegativeSurface {

    () log:onNegativeSurface {
        _:A a :Journal .
    } .
    
    ( ) log:onNegativeSurface {
        :Surface :is :JournalLess .
    } .
} .
\end{lstlisting}

Explicit negation is also distinct from SNAF, which is available in Notation3 by the \texttt{log:notIncludes} builtin predicate\footnote{\url{https://w3c.github.io/N3/reports/20230703/builtins.html\#log:notIncludes}}. Both NAF and SNAF styles of negation make statements about information that cannot be derived from a knowledge base. NAF requires assumptions about the rest of the world, thus implying complete knowledge about missing information. SNAF requires assumptions about missing triples within the boundaries of an RDF document. In contrast, explicit negation does not require any assumptions about missing information. Instead, RDF Surfaces can explicitly state negative information or derive explicit negative information. 

\subsection{Disjunctions in the data, antecedent and consequent of an implication}\label{s5.4}

Following the hints in \cref{s4.4}, a disjunction can be constructed using a combination of negations with conjunctions given that $A \lor B \equiv \lnot ( \lnot A \land \lnot B )$. \cref{list:disjunction_data} provides a fictional example of such disjunction by adding nested negative surfaces in a negative surface. Expressed in a symbolic form this becomes:

\begin{equation*}
    \forall x \exists y : \langle\texttt{x a :JournalArticle}\rangle \lor 
   \left( \langle\texttt{x a :Preprint}\rangle  \land \langle\texttt{y :reviewed x}\rangle \right) 
\end{equation*}

which is equivalent to:

\begin{equation*}
  \lnot \left( \exists x : \lnot \langle \texttt{x a :JournalArticle} \rangle \land 
     \lnot \left( 
      \exists y : \langle\texttt{x a :Preprint}\rangle  \land \langle\texttt{y :reviewed x}\rangle
     \right)
   \right).
\end{equation*}

\begin{lstlisting}[numbers=left,label=list:disjunction_data,caption={Adding nested negative surfaces in a negative surface creates a disjunction. The RDF Surfaces graph means, "Any article is a journal article or a preprint, and some S reviewed it, or both." }]
@prefix log: <http://www.w3.org/2000/10/swap/log#> .
@prefix : <https://example.org/ns#> .

( _:X ) log:onNegativeSurface {

    ( ) log:onNegativeSurface {
         _:X a :JournalArticle .
    } .
    
    ( _:Y ) log:onNegativeSurface {
        _:X a :Preprint .
        _:Y :reviewed _:X.    
    } .
} .
\end{lstlisting}

\cref{list:disjunction_data} can be read in many ways, all equally valid, and all share the same meaning:

\begin{itemize}
 \item Everything is a journal article or a preprint that was also reviewed by someone or both.
 \item If there is something that is not a journal article, it is a preprint, and someone reviewed it.
 \item If there is something that is not a preprint that someone reviewed, then it is a journal article.
\end{itemize}

This is because $\lnot X \rightarrow Y \equiv \lnot Y \rightarrow X \equiv X \lor Y$. RDF Surfaces makes the equivalence of all these readings explicit. 

To illustrate the effect of disjunction in reasoning, a negation can be added on the default positive surface, which, for instance, expresses that \texttt{$\langle$:MyArticle a :JournalArticle$\rangle$} is "false":

\begin{lstlisting}[numbers=none]
@prefix log: <http://www.w3.org/2000/10/swap/log#> .
@prefix : <https://example.org/ns#> .

( ) log:onNegativeSurface {
    :MyArticle a :JournalArticle .
} .
\end{lstlisting}

Applying the diagram rules of \cref{s4.4} the following deduction is possible:

\begin{itemize}
 \item \texttt{() log:onNegativeSurface \{ \_:X a :JournalArticle \}} can be deiterated from \cref{list:disjunction_data}  using rule \textbf{R4} when \texttt{\_:X} is equal to \texttt{:MyArticle}.
 \item This leaves us with a double cut that can be removed using rule \textbf{R3}.
\end{itemize}

\begin{figure}[hbtp]
\centering
\includegraphics[width=310pt]{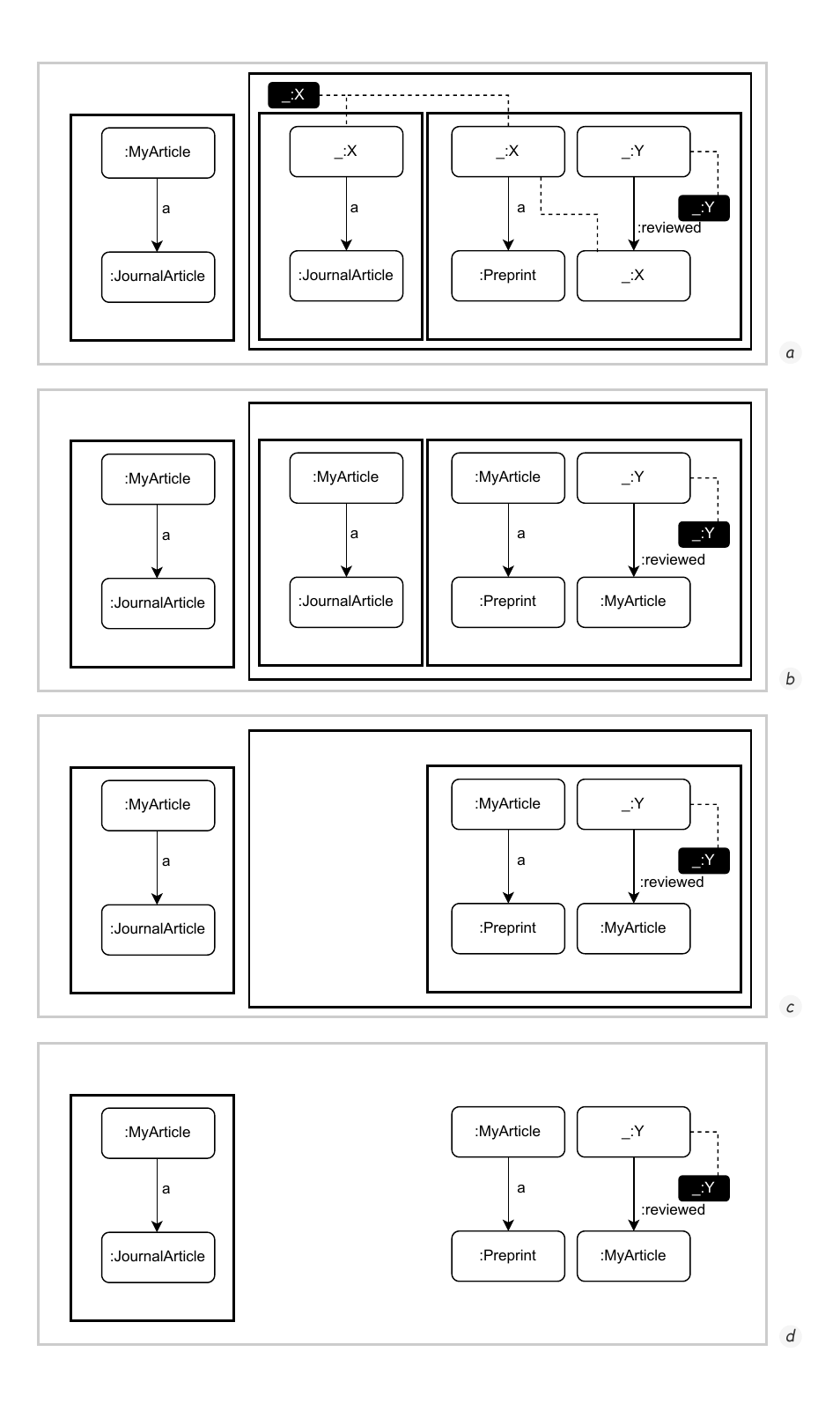}
\caption{(a) \cref{list:disjunction_data} with the added \texttt{$\lnot \langle$:MyArticle a :JournalArticle$\rangle$} on the default (positive) surface; (b) deiteration of graffiti nodes by setting \texttt{\_:X} equal to \texttt{:MyArticle}; (c) deiteration using rule \textbf{R4}; (d) removing the double cut using rule \textbf{R3}.}
\label{fig:s5_multiple_readings}
\end{figure}

As result, the deductive closure of \cref{list:disjunction_data} contains the RDF triples:

\begin{lstlisting}[numbers=none]
@prefix : <https://example.org/ns#> .

:MyArticle a :Preprint .
_:Y :reviewed :MyArticle .
\end{lstlisting} 

The diagram form of \cref{list:disjunction_data} and the diagrammatic derivation is available in \cref{fig:s5_multiple_readings}.

Disjunctions are not only available in the data but can also be added to the antecedent and consequent of an implication. That is, both forms are possible:

\begin{equation*}
 \forall x : \left( \langle\texttt{x a  :Journal}\rangle \lor  \langle\texttt{x a  :Book}\rangle \right) \rightarrow \langle \texttt{x a  :Publication} \rangle 
\end{equation*}

\noindent and

\begin{equation*}
 \forall x :   \langle\texttt{x a  :Publication}\rangle  \rightarrow \left(  \langle \texttt{x a  :Journal} \rangle \lor \ \langle \texttt{x a  :Book} \rangle \right).
\end{equation*}

Disjunctions in the consequent is a feature in RDF Surfaces unavailable in N3Logic. The latter implication could be written in RDF Surfaces using \cref{list:disjunction_consequent}.

\begin{lstlisting}[numbers=left,label=list:disjunction_consequent,caption={An RDF Surfaces graph with a disjunction in the conclusion with meaning: "Anything that is a publication is a journal or a book, or both."}]
@prefix log: <http://www.w3.org/2000/10/swap/log#> .
@prefix : <https://example.org/ns#> .

( _:X ) log:onNegativeSurface {

    _:X a :Publication .
    
    ( ) log:onNegativeSurface {
        ( ) log:onNegativeSurface {

            ( ) log:onNegativeSurface {
                 _:X a :Journal .
            } .
    
            ( ) log:onNegativeSurface {
                 _:X a :Book .
            } .
        } .
     } .
} .
\end{lstlisting}

Thus, complex implications can be composed using a copy-and-paste approach with simpler constructs. The pattern of nested negative surfaces on lines 4-8 starts an implication, and the pattern of negative surfaces on lines 9-18 is that of a disjunction.


\section{RDF Surfaces reasoner implementation}\label{s6}

While Peirce's diagram rules provide one approach to the logic inferencing of RDF Surfaces, it is not the sole method available. We use Peirce's diagram rules in this paper to convince the reader of the application of FOL using RDF Surfaces. Using four relatively simple rules, authors of RDF Surfaces can make simple derivations and check the consequences of surface logic by hand. However, this does not imply that automated systems must use this derivation method. The world of automatic theorem proving is extensive, featuring many FOL provers and an active community including a "World Championship for Automated Theorem Proving" (CADE).\footnote{\url{https://tptp.org/CASC/}} 

We are currently experimenting with four implementations specially targeted to RDF Surfaces. Only one implementation uses a direct translation of Peirce diagram rules in its codebase:

\begin{itemize}
    \item \textit{EYE}~\cite{eye} is implemented in SWI-Prolog\footnote{\url{https://www.swi-prolog.org}} and is based on forward and backward chaining. The resolution algorithm rewrites RDF Surfaces into a conjunction of disjunctive normal forms (DNF). The codebase was originally written to support the syntax and semantics of Notation3 but also implements RDF Surfaces semantics.
    \item \textit{Retina}~\cite{retina} is also implemented in Prolog and can be run in Trealla\footnote{https://github.com/trealla-prolog/trealla} and Scryer\footnote{\url{https://www.scryer.pl}}. The codebase is a rewrite of EYE targeted to processing RDF Surfaces.
    \item \textit{Latar}~\cite{latar} uses a calculus that is directory inspired by Beta graph reasoning implemented in SWI-Prolog.
    \item \textit{Tension.js}~\cite{tension} is implemented in Typescript and uses a similar resolution algorithm as EYE.
\end{itemize}

Of all these implementations, EYE is the most mature and was used extensively in our experiments to apply RDF Surfaces to real-world use cases. EYE provides a command line and a browser version.  At \url{https://w3c-cg.github.io/rdfsurfaces/demonstrator/}, an experimental RDF Surfaces implementation using EYE is available. The following RDF Surfaces features are already available in EYE:

\begin{itemize}
   \item Full support of the RDF Surfaces syntax as presented in this paper.
   \item Explicit scoping of logical variables.
   \item Existential closure of free variables on the default (positive) surface.
   \item Explicit negation of triples and conjunctions of triples using the \texttt{log:onNegativeSurface} predicate.
   \item Disjunctions in the data, antecedent and consequent of an implementation.
\end{itemize}

In the current version, EYE v10.10.0, there are some known limitations. Due to the extensive support for forward and backward chaining, our focus was mainly on use cases with an implication structure. Additionally, every implementation is expected to have computational limits due to the undecidable nature of the underlying logic. It is possible to create formulas that never halt. Creating formulas that will result in an incomplete answer is possible. 

EYE, building on its foundations in Notation3 reasoning, also offers several extensions that are not part of the core RDF Surfaces, including functional predicates for list, string, and mathematical calculations.

In EYE and all other RDF Surfaces implementations, there is no distinction between assertion and theory boxes, which are typically required in many knowledge processing systems. RDF Surfaces transport data and logic using a common embedded RDF syntax; all assertions and theory statements can be combined in one document. 

Two methods are available for querying data in an RDF Surfaces graph in all implementations. 

\textit{Proof by contradiction} checks if a graph $G$ is available in the knowledge base by adding the negated $\lnot G$ to the knowledge base and tests if this leads to a contradiction. 

\textit{Proof by negation} checks if a graph $\lnot G$ is available in the knowledge base by adding negated $\lnot\lnot G \equiv G$ to the knowledge base and tests if this leads to a contradiction. 

A contradiction in EYE is implemented by an "inference fuse." The EYE reasoner will stop running when such an "inference fuse" is detected and will display the context in which it happened.

Both proof methods are based on classical logic but are quite limited. They only provide yes/no answers: is a graph part of a knowledge base (and its derivations) or not? To answer a more generic query, "show me all triples and derived triples that match a triple/graph pattern," an experimental new surface was added to EYE: the query surface \texttt{log:onQuerySurface}. To each RDF Surfaces document, one or more query surfaces can be appended to inspect which bindings for graffiti nodes are available.
\Cref{list:query_surface} provides an example of a query surface that asks which patterns can be found in a knowledge base for accreditations. 

\begin{lstlisting}[numbers=left,label=list:query_surface,caption={An example query surface to retrieve all triples and derived triples that match a triple pattern.}]
(_:S _:O) log:onQuerySurface {
    _:S :accredit _:O .
} .
\end{lstlisting}

Applied to \cref{list:graffiti_rdf} this query surface should result in the following triples on the standard output:

\begin{lstlisting}[numbers=none]
_:e1 :accredit :JournalA .
\end{lstlisting}

Proof by contradiction and proof by negation are common derivation techniques. However, contradictions can appear in RDF Surfaces without adding a negated query. The Web can and will be contradictory. Web agents require an approach to address this challenge. Due to the principle of explosion, anything can be proven from a contradiction. If a knowledge base leads to a contradiction, something must be wrong for bad and good reasons. A bad reason includes, for instance, errors in a knowledge base that require correction or data from unreliable sources that should be disregarded. A good reason could include genuine conflicts in human knowledge\footnote{These days, we could add machine-generated knowledge when addressing this issue.}, which indicate starting points for research and academic discourse. In both cases, human intervention or heuristics based on some oracle could decide which data to include in a derivation and how to resolve conflicts. In our opinion, these explicit contradictions are safer on an open Web -- in some way even desired -- rather than implicit assumptions about negative information with possible contradictions that cannot be discovered.


\section{Examples}\label{s7}

With the help of the results of \cref{s5} and \cref{s4}, we are now ready to apply them to the use cases presented in \cref{s2}.

\subsection{Scholarly communication}\label{s7.1}

For the scholarly communication use case, we envision a scholarly network of researchers who share their preferences when searching for publication venues. These preferences are preference documents that are serialized as N3S. These policies can contain information on what preference is regarded as positive, what certainly is not the case or preferences that are only "true" under some conditions. The latter takes the form of an implication. \cref{s2.1} provides an example of such a researcher preference:

\begin{itemize}
 \item \textit{Researcher X preferences}: Researcher X prefers preprints in a subject repository, a journal that does not charge APC costs, or a journal indexed in WOS.
\end{itemize}

One way to express this preference in FOL is by using disjunctions in the antecedent of an implication, as demonstrated in \cref{eq:6_schol_comm_1}.

\begin{equation}
 \label{eq:6_schol_comm_1}
 \begin{split}
    \forall \texttt{x} : & \Big( \langle\texttt{x a :SubjectRepository}\rangle  \  \lor \\
     &( \langle\texttt{x a :Journal}\rangle \land \lnot \langle\texttt{x :charges :APC}\rangle )  \ \lor \\
     & ( \langle\texttt{x a :Journal}\rangle \land \langle\texttt{:WOS :indexed x}\rangle  ) \Big) \\
     &\longrightarrow  \langle\texttt{x a :ResearcherPreference}\rangle .
  \end{split}
\end{equation}

For the preference \cref{eq:6_schol_comm_1}, the universal quantified variable $x$ stands for a publication venue. When the conditions hold for $x$, it is a preference for a particular researcher.

An alternative way to formulate this preference with the same meaning can be seen in \cref{eq:6_schol_comm_2}.

\begin{equation}
 \label{eq:6_schol_comm_2}
 \begin{split}
     \Big( \forall \texttt{x} :  \langle\texttt{x a :SubjectRepository}\rangle  \rightarrow  \langle\texttt{x a :ResearcherPreference}\rangle  \Big) \ \land \\
     \Big( \forall \texttt{x} :  ( \langle\texttt{x a :Journal}\rangle \land \lnot \langle\texttt{x :charges :APC}\rangle )  \rightarrow  \langle\texttt{x a :ResearcherPreference}\rangle  \Big) \ \land \\
     \Big( \forall \texttt{x} :  ( \langle\texttt{x a :Journal}\rangle \land \langle\texttt{:WOS :indexed x}\rangle  )  \rightarrow  \langle\texttt{x a :ResearcherPreference}\rangle  \Big) . 
  \end{split}
\end{equation}

We made here use of the fact that $(A \rightarrow C) \land (B \rightarrow C) \equiv ( A \lor B ) \rightarrow C$ which can be easily derived using Peirce's Alpha system\footnote{
	The derivation starts on the left with the diagram for $(A \rightarrow C) \land (B \rightarrow C)$. Using diagram rule \textbf{R3}, a double negated surface can be drawn around this diagram. Using diagram rule \textbf{R1}, a new \fbox{C} symbol can be added on the outer negative surfaces with parity 1. With diagram rule \textbf{R4} this \fbox{C} can be deiterated from the inner nested surfaces with nesting level 3. The result is the diagram for $( A \lor B ) \rightarrow C$.\\
	\includegraphics[width = 300pt]{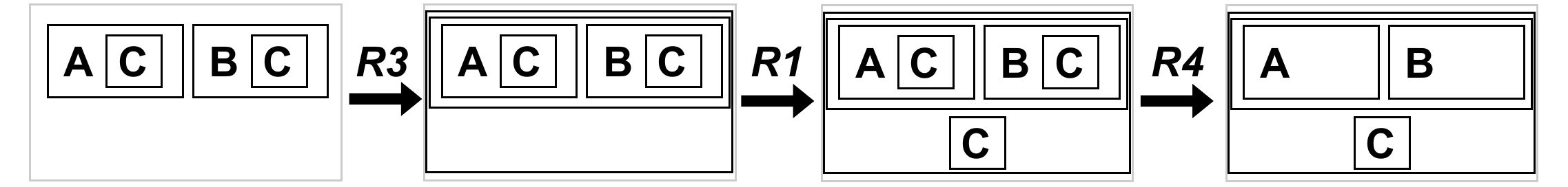}
}. This alternative formulation has the advantage that formula  \cref{eq:6_schol_comm_2} as an RDF Surfaces serialization does not require the double nesting of all disjunction elements. New preference options can be added to the RDF Surfaces preferences document by appending a new implication pattern rather than inserting a new RDF Surfaces statement into an existing disjunction. The translation of the preferences of researcher X, as stated in \cref{eq:6_schol_comm_2}, into RDF Surfaces is provided in \cref{list:schol_comm_1}.

\begin{lstlisting}[numbers=none,label=list:schol_comm_1,caption={The translation of the symbolic \cref{eq:6_schol_comm_2} into RDF Surfaces N3S.}]
@prefix : <https://example.org/ns#> .
@prefix log: <http://www.w3.org/2000/10/swap/log#> .

# Pref 1 . Publications in a subject repo
(_:X) log:onNegativeSurface {
  _:X a :SubjectRepository .

  () log:onNegativeSurface {
    _:X a :ResearcherPreference.
  } . 
} .

# Pref 2 . Publications by a journal that doesn't charge APC costs
(_:X) log:onNegativeSurface {
    _:X a :Journal .

    () log:onNegativeSurface {
        _:X :charges :APC .
    } .

    () log:onNegativeSurface {
        _:X a :ResearcherPreference.
    } .
} .

# Pref 3 . Publications by a publisher that is in WOS
(_:X) log:onNegativeSurface {
   _:X a :Journal .
   :WOS :indexed _:X .

   () log:onNegativeSurface {
        _:X a :ResearcherPreference.
   } .
} .
\end{lstlisting}

The same procedure can be done for the departmental Y preferences that were stated as follows:

\begin{itemize}
 \item \textit{Department Y preferences}: All publications must be journals that are indexed in WOS.
\end{itemize}

A direct translation into FOL follows in  \cref{eq:6_schol_comm_3}.

\begin{equation}
 \label{eq:6_schol_comm_3}
    \forall \texttt{x} :  ( \langle\texttt{x a :Journal}\rangle \land \langle\texttt{:WOS :indexed x}\rangle ) \rightarrow \langle\texttt{x a :DepartmentPreference}\rangle  )
\end{equation}

The RDF Surfaces version of \cref{eq:6_schol_comm_3} is available \cref{list:schol_comm_2}.

\begin{lstlisting}[numbers=none,label=list:schol_comm_2,caption={The translation of the symbolic \cref{eq:6_schol_comm_3} into RDF Surfaces N3S.}]
@prefix : <https://example.org/ns#> .
@prefix log: <http://www.w3.org/2000/10/swap/log#> .

# Pref 1 . Only journals that are indexed in WOS
(_:X) log:onNegativeSurface {
   _:X a :Journal .
   :WOS :indexed _:X .

   () log:onNegativeSurface {
      _:X a :DepartmentPreference .
   } .
} .
\end{lstlisting}

The policies for publishing venues can be spread around the Web. Sources that describe journal information can include information such as the journal homepage, the publisher's website, the library database, and indexing services, such as Web of Science, to name a few. The nature of this information is inherently decentralized and, when provided as Linked Data, distributed using many ontologies. Establishing a common method for creating negative facts becomes essential in such an environment. For instance, it is quite common for journals to charge APC costs. That a particular journal does not charge APC costs is a negative fact that benefits the budgets of many researchers. In \cref{eq:6_schol_comm_4}, we symbolize a summary of publishing venue facts that state:

\begin{itemize}
 \item ABC, DEF, and GHI are journals.
 \item JKL is a subject repository.
 \item ABC charges APC costs, but GHI does not charge these costs.
 \item ABC and DEF are indexed in the WOS database, but GHI is not.
\end{itemize}

\begin{equation}
 \label{eq:6_schol_comm_4}
  \begin{split}
   &\langle\texttt{:ABC a :Journal}\rangle \land  \langle\texttt{:DEF a :Journal}\rangle \land  \langle\texttt{:GHI a :Journal}\rangle \ \land \\
   &\langle\texttt{:JKL a :SubjectRepository}\rangle \  \land \\
   &\langle\texttt{:ABC :charges :APC}\rangle \land \lnot  \langle\texttt{:GHI :charges :APC}\rangle \ \land\\
   &\langle\texttt{:WOS :indexed :ABC}\rangle \land  \langle\texttt{:WOS :indexed :DEF}\rangle \land \lnot\langle\texttt{:WOS :indexed :GHI}\rangle
  \end{split}
\end{equation}

The translation of \cref{eq:6_schol_comm_4} into RDF Surfaces is provided in \cref{list:schol_comm_4}.

\begin{lstlisting}[numbers=none,label=list:schol_comm_4,caption={The translation of the symbolic \cref{eq:6_schol_comm_4} into RDF Surfaces N3S.}]
@prefix : <https://example.org/ns#> .
@prefix log: <http://www.w3.org/2000/10/swap/log#> .

## Journal facts
:ABC a :Journal .
:DEF a :Journal .
:GHI a :Journal .

# APC facts
:ABC :charges :APC .

## GHI is a journal that does not require APC costs
() log:onNegativeSurface {
    :GHI :charges :APC .
} .

# WOS Facts
:WOS :indexed :ABC , :DEF .

() log:onNegativeSurface {
    :WOS :indexed :GHI .
} .

## JKL is a subject repository
:JKL a :SubjectRepository .
\end{lstlisting}

In our vision, facts about publishing venues are publicly shared. This allows Web agents to harvest and match the data against researcher and department policies. Consequently, this process can be used to make informed publishing recommendations as a service. For instance, when a Web agent gets hold of the policies expressed in \cref{eq:6_schol_comm_1}, \cref{eq:6_schol_comm_2}, \cref{eq:6_schol_comm_3}, and \cref{eq:6_schol_comm_4}, a logical query can be posed that asks if journal ABC is a researcher and department preference, by adding the negation of $ \langle\texttt{:ABC a :ResearcherPreference}\rangle \land  \langle\texttt{:ABC a :DepartmentPreference}\rangle$ to the knowledge base and test if this leads to a contradiction. 

At \url{https://w3c-cg.github.io/rdfsurfaces/demonstrator/}, an experimental RDF Surfaces implementation using EYE is available to test the examples provided in this section. The RDF Surfaces from this section can be consulted as one resource at \url{https://w3c-cg.github.io/rdfsurfaces/examples/scholary\_publication.n3}.

For illustration purposes, we assume a Web agent wants to consult these preferences and adds a negated query at the end:

\begin{lstlisting}[numbers=none,label=list:negation_query]
() log:onNegativeSurface {
    :ABC a :DepartmentPreference .
    :ABC a :ResearcherPreference .
} .
\end{lstlisting}

The negation: 

\begin{equation*}
 \lnot ( \langle\texttt{:ABC a :ResearcherPreference}\rangle \land  \langle\texttt{:ABC a :DepartmentPreference}\rangle )
\end{equation*}
 
\noindent leads to a contradiction. Therefore, the Web agent can conclude that journal ABC is a researcher and department preference.

\begin{figure}[t]
\centering
\includegraphics[width = 250pt]{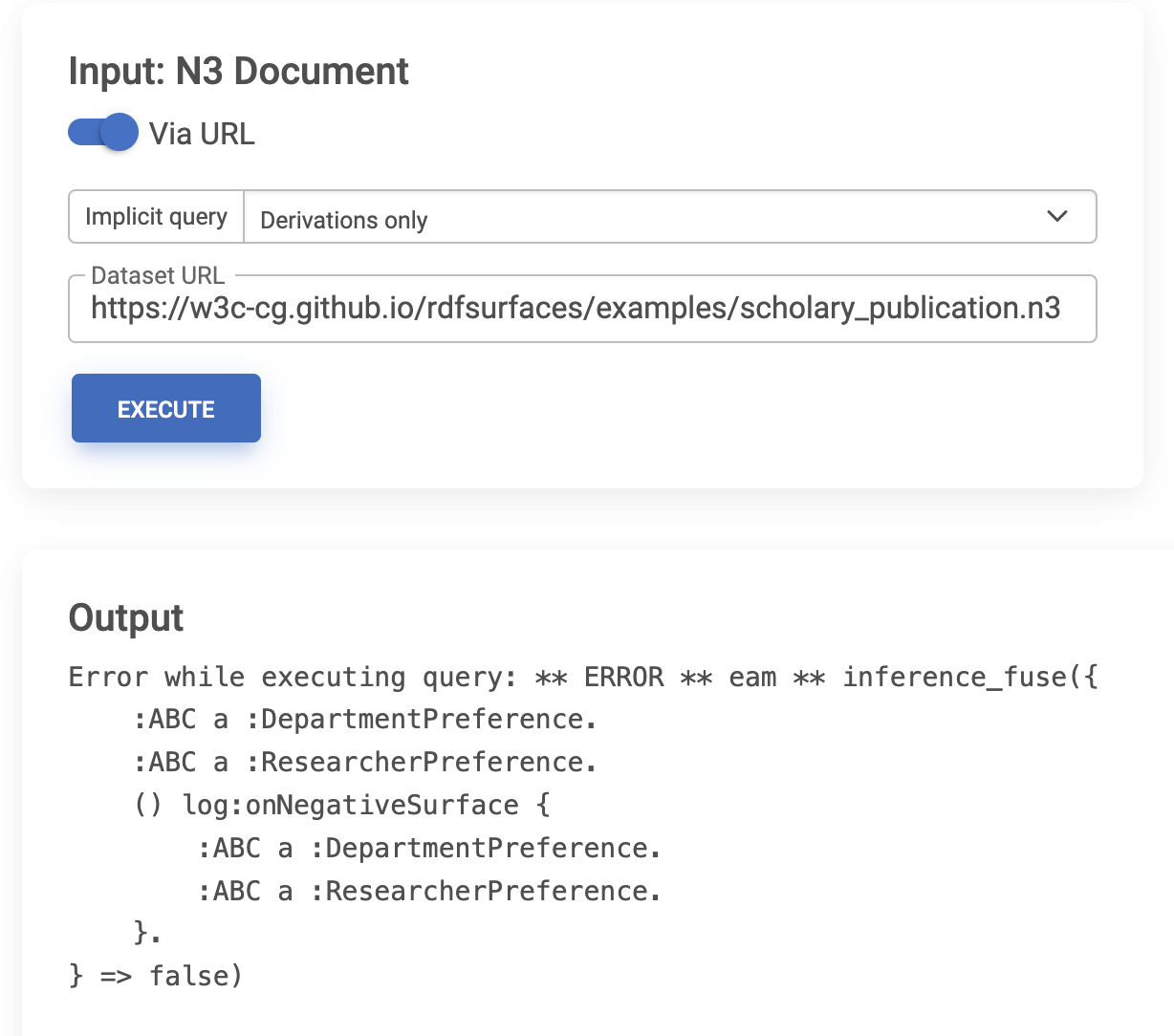}
\caption{If adding negated triples to an RDF Surfaces document results in a contradiction (expressed as an inference fuse in EYE), the positive version of the triples can be asserted instead. In this example $\langle\texttt{:ABC a :ResearcherPreference}\rangle$ and $\langle\texttt{:ABC a :DepartmentPreference}\rangle$ is part of the deductive closure.}
\label{fig:s6_scholarly_publishing}
\end{figure}

\cref{fig:s6_scholarly_publishing} illustrates this example where, after specifying the location of the RDF Surfaces knowledge base plus the negated query, the derivation leads to an "inference fuse", indicating a contradiction.  

\subsection{Medicine Prescription}\label{s7.2}

In the healthcare domain, we envision knowledge systems that include both positive and negative information about medications and policies on how these medications can be prescribed to patients with pre-existing conditions. These systems would consider that for example certain medications may cause allergic reactions when prescribed to patients. Consider a collection of medicines for a medical prescription use case: high-dosage aspirin, low-dosage aspirin, and beta-blockers. A high dosage of aspirin is an effective treatment for a fever. In turn, a low dosage of aspirin or a prescription of beta-blockers are both considered to be an effective treatment for acute myocardial infarction. However, both high and low dosages of aspirin may only be prescribed when a patient is known not to be allergic to aspirin. Aspirin can also not be prescribed when a patient has active peptic ulcer disease. Beta-blockers, on the other hand, cannot be prescribed when a patient suffers from severe asthma.

We assume the following expert policies in terms of medicine prescription:

\begin{itemize}
    \item A low aspirin dosage can be prescribed for a patient with a fever and with neither an allergy to aspirin nor active peptic ulcer disease.
    \item For a patient with acute myocardial infarction but without an allergy to aspirin or active peptic ulcer disease, a high dosage of aspirin can be prescribed.
    \item For a patient with acute myocardial infarction but without severe asthma or chronic obstructive pulmonary disease, beta-blockers can be prescribed.
\end{itemize}

The first policy indicates that if a given patient has a fever (A), one of the following must hold: the patient has an aspirin allergy (B), has active peptic ulcer disease (C), or is prescribed a high dosage of aspirin (D). This can be inferred from the logical equivalence $\neg (A \land \neg B \land \neg C \land \neg D ) \equiv (A \land \lnot B \land \lnot C ) \rightarrow D \equiv A \rightarrow (B \lor C \lor D )$. In other words, the Peircian diagram that corresponds with these formulas has equal readings\footnote{
	Diagram (a) can be read in many ways: $\neg (A \land \neg B \land \neg C \land \neg D )$, $(A \land \lnot B \land \lnot C ) \rightarrow D$, $(\lnot B \land \lnot C \land \not D) \rightarrow \lnot A$, $(A \land \lnot C \land \lnot D) \rightarrow \lnot B$, etc. Diagram (b) can be read as $A \rightarrow (B \lor C \lor D)$ and is diagram (a) with an added double negative surface around the negated $B$, $C$, and $D$ which can be removed using diagram rule \textbf{R3}. \\
	\includegraphics[width = 150pt]{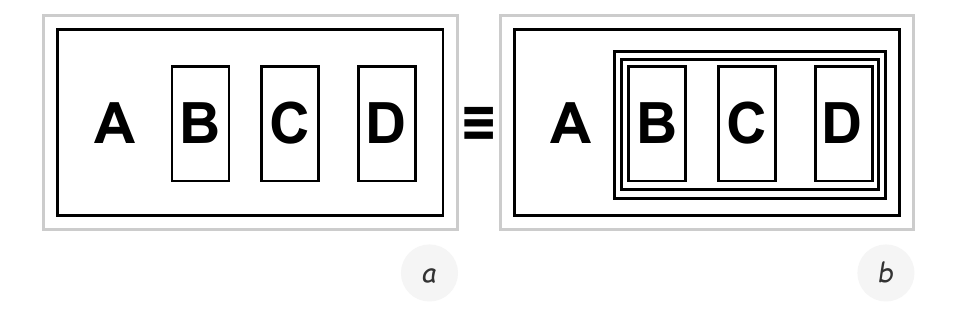}
}. Any of these three forms could be used to translate the policy in a FOL format.  In \cref{eq:med_prescrib}, we use a symbolic form that closely matches the method applied in \cref{eq:6_schol_comm_2} for researcher preferences.

\begin{equation}
   \label{eq:med_prescrib}
   \begin{split}
     \forall \texttt{x} : \Big(
       &\langle \texttt{x :has :Fever}\rangle \ \land \\
       &\lnot \langle \texttt{x :has :AllergyForAspirin} \rangle \ \land \\
       &\lnot \langle \texttt{x :has :ActivePepticUlcerDisease} \rangle \ \Big) \\
       &\longrightarrow \langle \texttt{x :isPrescribed :aspirinHighDose} \rangle 
   \end{split}
\end{equation}

The translation for the other two pieces of expert knowledge can be constructed analogously. The RDF Surfaces version of these policies is provided in Appendix A. 

With the expert knowledge encoded in RDF Surfaces, we are able to infer the correct prescription for a patient, given their medical condition as well as allergies (or lack thereof). We consider the following two patients, Ann and Joe. Ann has a fever and does not have an allergy to aspirin or an active peptic ulcer disease. Joe suffers from acute myocardial infarction and is allergic to aspirin. However, he does not have active peptic ulcer disease, severe asthma, or chronic obstructive pulmonary disease. These facts can be translated into symbolic form as demonstrated by \cref{eq:patient1_info} and \cref{eq:patient2_info} with a translation in RDF Surfaces in Appendices B and C.

\begin{equation}
 \label{eq:patient1_info}
 \begin{split}
 &\langle \texttt{:Ann :has :Fever} \rangle \ \land \\
 & \lnot  \langle \texttt{:Ann :has :AllergyForAspirin}  \rangle \ \land \\
 & \lnot   \langle \texttt{:Ann :has :ActivePepticUlcerDisease} \rangle
 \end{split} 
\end{equation}

\begin{equation}
 \label{eq:patient2_info}
 \begin{split}
 &\langle \texttt{:Joe :has :AcuteMyocardialInfarction} \rangle \ \land \\
 &\langle \texttt{:Joe :has :AllergyForAspirin} \rangle \ \land \\
 &\lnot  \langle \texttt{:Joe :has :ActivePepticUlcerDisease}  \rangle \ \land \\
 &\lnot  \langle \texttt{:Joe :has :SevereAsthma}  \rangle \ \land \\
 &\lnot   \langle \texttt{:Joe :has :ChronicObstructivePulmonaryDisease} \rangle
 \end{split} 
\end{equation}

The RDF Surfaces from Appendices A, B, and C can be consulted as one resource at \url{https://w3c-cg.github.io/rdfsurfaces/examples/medication\_prescription.n3}. Using the experimental RDF Surfaces implementation of \cref{s6}, this resource can be consulted to test if we can prescribe a high-dose aspirin for patient Ann and a beta blocker for patient Joe. The procedure is similar to the one in the previous section. We can add to the stated RDF Surfaces a negative query and check for a contradiction:

\begin{lstlisting}[numbers=none,label=list:patient_query_1]
() log:onNegativeSurface {
    :Ann :isPrescribed :aspirinHighDose .
    :Joe :isPrescribed :betaBlocker .
} .
\end{lstlisting}

\begin{figure}[t]
\centering
\includegraphics[width = 250pt]{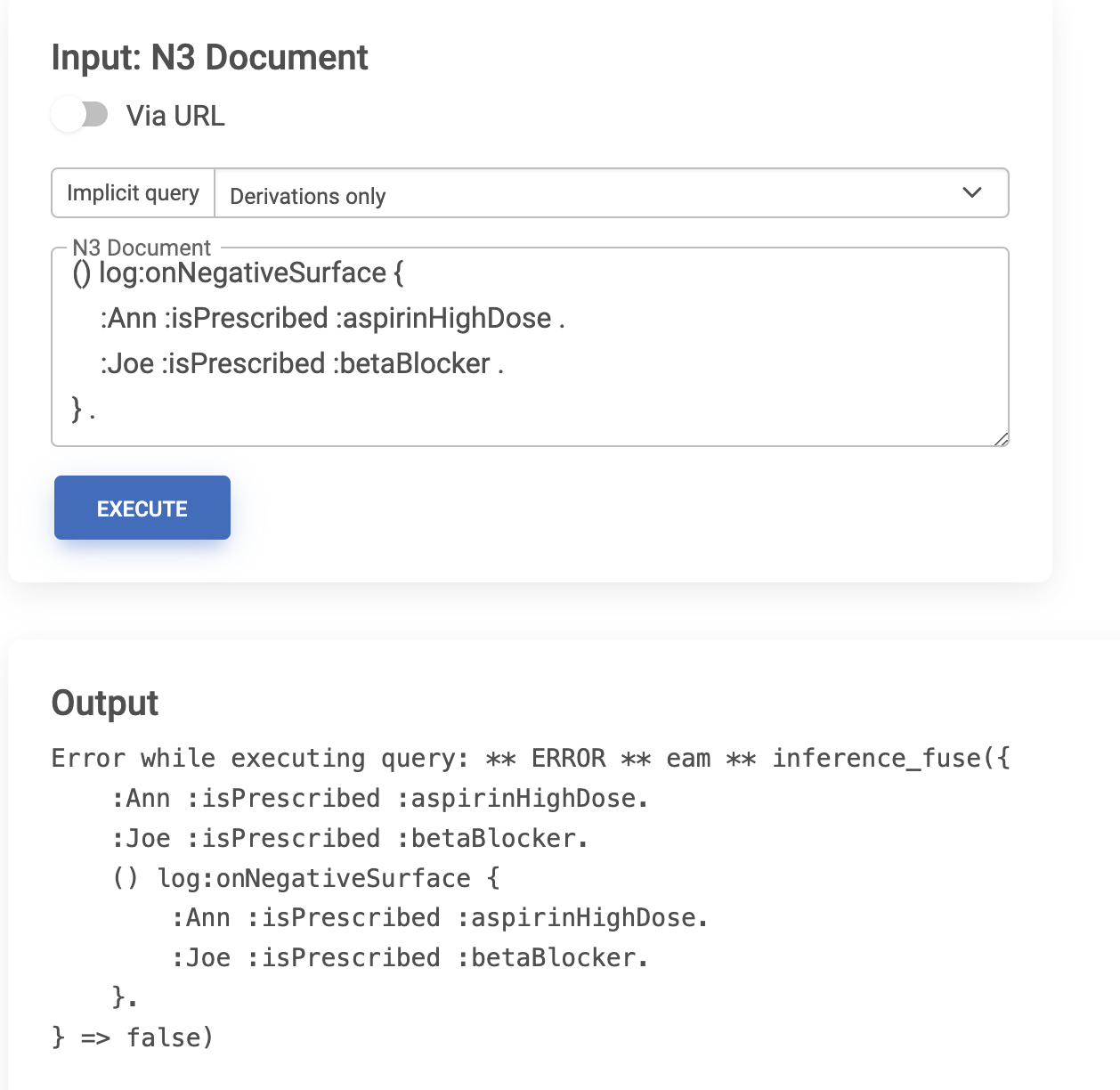}
\caption{Addition of a negated conjunction to the healthcare knowledge base plus patient data leads to a contradiction. Therefore, the positive version of the conjunction can be asserted.}
\label{fig:s6_health_care}
\end{figure}

\begin{figure}[t]
\centering
\includegraphics[width = 250pt]{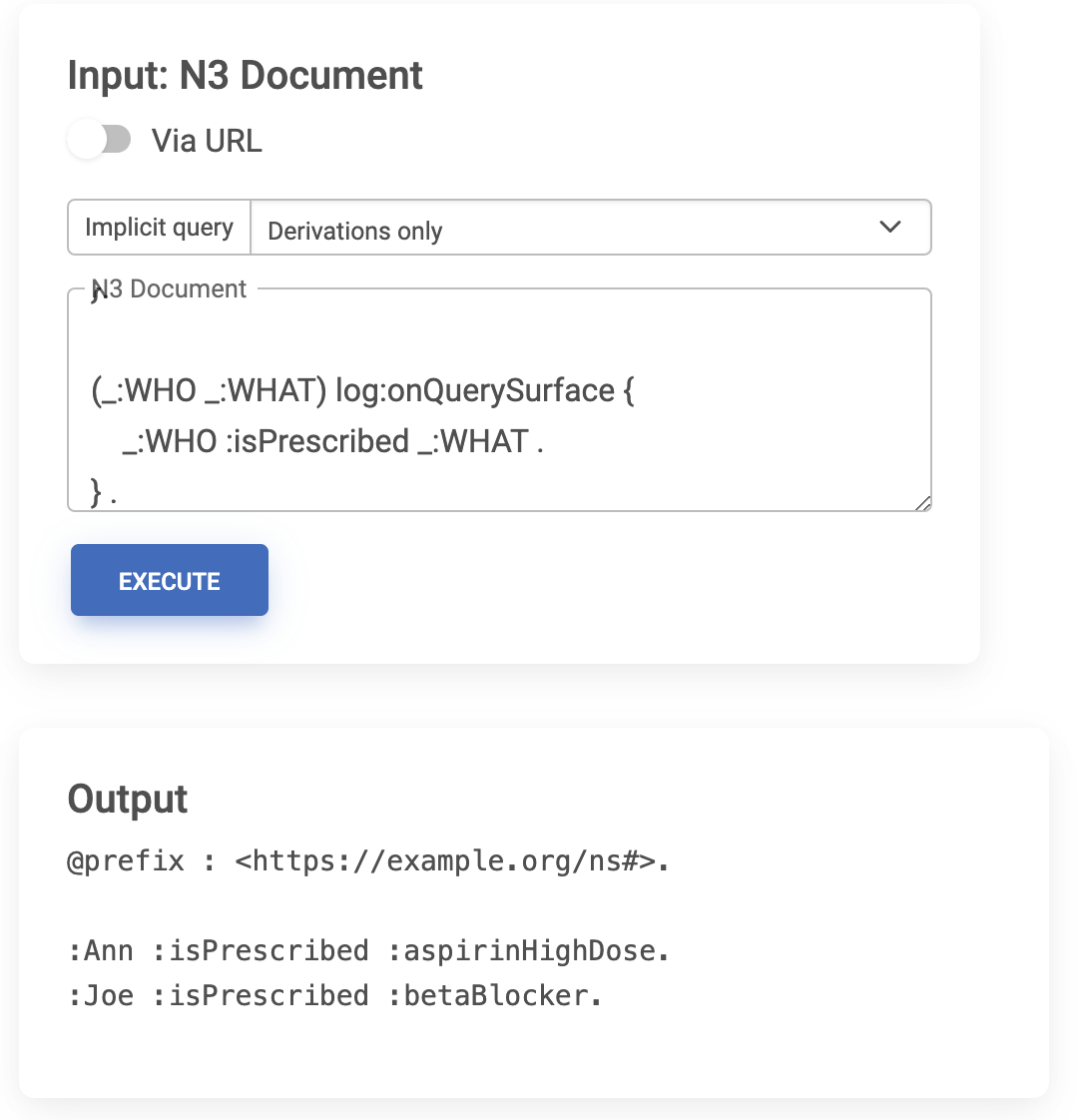}
\caption{The addition of a query surface to an RDF Surfaces document can be used to bind all graffiti nodes and inspect all triples that can be derived.}
\label{fig:s6_health_care_2}
\end{figure}

\cref{fig:s6_health_care} illustrates this example where, after copy and pasting the RDF Surfaces knowledge base and adding the negative query, the derivation leads to an "inference fuse", indicating a contradiction. The \textit{negation}:

\begin{equation*}
 \lnot ( \langle\texttt{:Ann :isPrescribed :aspirinHighDose}\rangle \land  \langle\texttt{:Joe :isPrescribed :betaBlocker}\rangle )
\end{equation*}
 
\noindent leads to a contradiction. Therefore, we can conclude that patient Ann can be prescribed a high aspirin dosage, and patient Joe can be prescribed a beta-blocker. 

Are these the only conclusions that can be drawn from this knowledge base and patient data? To know who can be prescribed what medication instead of a negative surface, the EYE query surface can be added to the knowledge base:

\begin{lstlisting}[numbers=none,label=list:patient_query_2]
(_:WHO _:WHAT) log:onQuerySurface {
    _:WHO :isPrescribed _:WHAT .
} .
\end{lstlisting}

When this query is run in the reasoning interface, as demonstrated in \cref{fig:s6_health_care_2}, the output will include:

\begin{lstlisting}[numbers=none,label=list:patient_query_result_2]
@prefix : <https://example.org/ns#>.

:Ann :isPrescribed :aspirinHighDose .
:Joe :isPrescribed :betaBlocker .
\end{lstlisting}

In this case the reasoner will find any possible binding in the deductive closure where

\begin{equation*}
\forall \texttt{x},\text{y} : \langle \texttt{x :isPrescribed y} \rangle 
\end{equation*}

leads to a contradiction. The query surfaces only includes RDF triples (not N3S) and in case of a contradiction it provides counter examples following this scheme.


\section{Discussion and future road map}\label{s8}

This paper discussed the need for an expressive Semantic Web Logic extending RDF. We especially identified classical negation as crucial for many use cases, including scholarly communication and healthcare. Therefore, we proposed RDF Surfaces, provided an abstract and a concrete syntax and first reasoner implementation, and showed how our use cases could benefit from it. The authors are currently experimenting with implementing RDF Surfaces in reasoners, and future experiments using existing FOL reasons from the TPTP project are on the near horizon. In 2022, a W3C Community group was formed to define the semantics and discuss the general requirements for implementations and Web logic.\footnote{\url{https://www.w3.org/community/rdfsurfaces/}}\textsuperscript{,}\footnote{\url{https://w3c-cg.github.io/rdfsurfaces/}} Of course, this new logic comes with many open challenges, and this vision paper is only the starting point of a longer endeavor. Below, we list the most important challenges ahead and briefly discuss possible ways to solve them.

\subsection{Negation on the Web}\label{s8.1}

With RDF Surfaces, we provided logic and syntax to express negative information on the Web using RDF. To implement negation, we extended the RDF model with surfaces that describe a (possibly empty) set of negated triples. These triples are quantified by adding graffiti nodes that act as existentially quantified variables. Combining these features with the default conjunction of triples under simple entailment semantics provides the full expressivity of FOL in RDF. Ambiguities in the quantification can be avoided by providing an explicit scope for graffiti nodes placed "on" a surface. We demonstrated how to solve real-world use cases in scholarly communication and healthcare that contain shared negative information, disjunctions, and implications in a decentralized setting. Using RDF Surfaces, Hayes's BLOGIC vision can be realized in concrete implementations, such as the EYE reasoner demonstrated in this paper.

The expressivity and portability of FOL provide the advantage that we can state what we want to say instead of only stating what machines can process. The Semantic Web is an endeavor to express the combined human knowledge irrespective of the computability of this knowledge. Consequently, we do not anticipate that any realistic Web reasoner will be capable of delivering a fully portable solution (P) and accepting every conceivable FOL input, consistently providing complete (C) and exhaustive responses to any query, and invariably terminating (H) within a finite time frame. Demanding these (P)+(C)+(H) attributes simultaneously is infeasible due to the undecidability of FOL and forms an iron triangle: at most, only two of these features can be selected for any implementation. If we want a portable Web logic, feature (P) is unavoidable. Any portable Web reasoner must choose to be complete (C) or always halt (H) in a finite time. In practice, this will lead to a necessary fragmentation and decentralization of the Web where many Web agents must compete to find inconsistencies or find all derivations that can be made from a decentralized knowledge graph. 

When a derivation or inconsistency is found, verifying that the given solution can be derived from the knowledge graph is possible. As Trakhtenbrot~\cite{trakhtenbrot_recursive_1953} demonstrated, this is a semi-decidable procedure\footnote{A semi-decidable procedure always halts and says "yes" when a statement $S$ validly derives a knowledge base $K$ but might not halt if $S$ is an invalid derivation.}. A more interesting solution would be publishing proof of every derivation. Rather than publishing what is a (derived) fact on the Web, Web agents could publish the proof of how they concluded these facts. For instance, in a scholarly communication setting, one would like to know why a journal was selected as a valid preference for a researcher. In a healthcare setting, one would trace why or why not a medication was provided to a patient.

We do not anticipate that every Web author should use FOL terminology to express knowledge and create logical formulas. We regard RDF Surfaces as a low-level language that can transport data and logic to the Web. Higher level vocabularies are still required to create a compact serialization for human consumption (human editors). The pragmatic choice will be to exchange knowledge using higher level vocabularies that can be "compiled" into RDF Surfaces logic. One such example is from the Flemish SHARCS project\footnote{\url{https://www.imec-int.com/en/research-portfolio/sharcs}} where a demonstrator was created to make ODRL policy actionable by compilation them to RDF Surfaces.\footnote{\url{https://github.com/eyereasoner/Notation3-By-Example/tree/main/examples/sharcs-odrl}} A similar approach using RDF Surfaces was taken by Zhuo and Zhuo (2024).

Extensions to RDF Surfaces would certainly be helpful when expressing non-monotonic negation. For instance, in the scholarly communication use case, expressing which topics are not part of a database is not always feasible. Notation3 provides the capabilities for scoped negation as failure that might need to be combined with RDF Surfaces. Real-world use cases also need to provide input on which extended predicates are required for things that are not easily expressable in FOL, such as calculations, string manipulation, and list manipulation. This requires a delicate balance between the expressiveness of the language for general computation use cases and the formal semantics of classic FOL.

\subsection{Syntax}\label{s8.2}

A first attempt at a hosting language for RDF Surfaces led us to use the Notation3 syntax in the form of "RDF Surfaces in N3" (N3S), which only requires a reduced set of features. Notation3 was only used for its syntactical features -- graph terms for creating a scope around negated triples and list terms for declaring graffiti nodes -- but not its semantic features. Some authors of this paper are active participants in both the Notation3 and RDF Surfaces groups, where overlap and differences between both approaches are debated. One point of argument is the scoping of blank nodes that in Notation3 semantics is local to a graph term but in RDF Surfaces explicit within a negative surface. Another point is Patel-Schneider's argument that a same-syntax extension of RDF to FOL necessarily leads to paradoxes due to self-referential statements~\cite{patel-schneider_rdf_2009}. We think we can 'wiggle out' self-referential statements due to introducing negative surfaces with a N3 list term in the subject position (where the graffiti nodes are declared). Using N3 list terms, negative surfaces inside an RDF Surfaces document do not have an identifier. Surfaces can be compared and be isomorphic but cannot be referred to. But this does not absolve us from paradoxes because these are very hard to avoid in highly expressive languages. At least, RDF Surfaces does not rely on potential self-referential statements.

\subsection{Formal semantics}\label{s8.3}

As RDF Surfaces is introduced as a logic, the obvious next step is to define it formally. We have already provided the syntax in this paper, but the semantics are only discussed informally and must be specified further. There are several possibilities to solve this issue. Given that Pat Hayes' original idea of a BLOGIC was inspired by existential graphs, we could base RDF Surfaces semantics on Beta graphs and map them to this framework. This would make it easy to define a calculus for RDF Surfaces by following existing literature, e.g., Zeman~\cite{zeman_graphical_1964} and Dau~\cite{dau}. A difficulty, however, is that Beta graphs do not have native support for constants. To formalize this, one would need to address this limitation, for example, by mimicking constants using relations in combination with existential quantification. Another possibility would be to base RDF Surfaces semantics on FOL, which is close to the approach we followed in this paper when we explained our examples.  An advantage of this approach would be that FOL is well understood regarding expressiveness and limitations. As RDF Surfaces extends RDF, RDF formalizations based on Beta graphs or FOL would have the burden of proof that the semantics is in line with the one of RDF. In the case of a FOL formalization, this could be done following the work of De Bruijn et al.~\cite{erdf}. This last aspect --  the fact that RDF Surfaces aim to extend RDF -- makes a formalization extending RDF semantics~\cite{hayes_rdf_2014} an interesting choice.  It would make sense only to consider simple entailment and D-entailment as the semantics or RDFS could be modeled through RDF Surfaces axioms. Of course, these axioms would need to be carefully designed, and the equivalence would need to be proven.

\subsection{Implementations}\label{s8.4}

To be of use for the Web, RDF Surfaces should not only be a theoretical framework, but should be applicable in practice. We need implementations that can check and exchange each other's findings to achieve this. We thus need to agree on the explicit syntax, and the N3S-based framework we introduced in this paper could be a starting point for that.  Reasoners should use this as input and apply a calculus that is proven correct and -- if possible -- also complete according to the semantics.  Currently, there are four experimental implementations based on our N3S-based syntax: Latar~\cite{latar}, whose calculus is directly inspired by Beta graph reasoning, EYE~\cite{eye}, which is originally a Notation3 reasoner but lately also provides support for RDF Surfaces, retina~\cite{retina} based on a rewrite of EYE for RDF Surfaces in Trealla and Scryer Prolog, and Tension.js~\cite{tension} a Typescript implementation. The calculus applied is close to FOL reasoning. To ensure interoperability between these reasoners and encourage more implementors to support RDF Surfaces, we need shared test cases that express the expected behavior of reasoning systems. A first repository providing such test cases is already available\footnote{\url{https://github.com/eyereasoner/rdfsurfaces-tests}}. Still, the test infrastructure needs to become more fine-grained to check for different types of problems like simple parsing errors or the correct application of single inference steps. Use cases like the ones we presented above will help generate more test cases and better understand the specific needs the reasoning should satisfy. They could help to decide on optimizations for data storage and inferencing.

\subsection{Relation to other Web logics and standards}\label{s8.5}

Alongside the formalization of the logic, its relationship to existing frameworks should be investigated. Since RDF Surfaces supports classical negation, existential quantification, and conjunction, it is likely that it can express FOL, though this claim must be proven. It is particularly interesting to explore how we can express OWL DL and its profiles~\cite{OWLprofiles} in RDF Surfaces, but other common Web frameworks should also be considered. 

The relationship to RDF Surfaces is clear for RDF and simple entailment, as the latter is defined as an extension of the former. However, the relationship is less clear for RDFS or rule formats like Notation3 Logic~\cite{n3spec} or SWRL~\cite{SWRL}. And it becomes even more difficult regarding WC3 recommendations like SHACL~\cite{Knublauch2017Shapes} or SPARQL~\cite{sparql}, which both support non-monotonic features. In its current form, as presented in this paper, RDF Surfaces cannot cope with non-monotonicity. Still, other aspects, like querying with recursive property paths, can be covered in RDF Surfaces.

\subsection{Extensions}\label{s8.6}
The last aspect mentioned, the current inability to support non-monotonic behavior, illustrates another direction for future research: it could be possible and maybe even necessary to support concepts, such as negation as failure or closed predicates~\cite{closed}, to be suitable to fully support use cases like the ones introduced in this paper. Notation3 and SPARQL come with built-in or filter predicates, which makes it easy to, e.g. perform operations on strings or sum up two integers. As RDF Surfaces is introduced as a logic facilitating practical use cases, it would be beneficial to include such predicates. As a third, but -- given the variety of use cases in the Web -- certainly not least extension, we believe that support for unasserted triples like it is proposed in RDF-star~\cite{rdf-star} and unasserted graphs as they exist in N3 would be beneficial for many use cases, like for example the exchange of provenance knowledge and the reasoning about it. To also support the latter, this support for unasserted knowledge should come with a mechanism that could assert it in the case of provenance, for example, if we decide to trust a source.

Of course, this section only provides a few examples of future development, and we hope our readers already have their own visions here. In that sense, we are very curious to see what the future brings, and we plan to work on these and other research to further RDF actively.

\section{Conclusion}\label{s9}

RDF Surfaces is a concrete implementation of Pat Hayes' BLOGIC vision of portable Web logic based on FOL. In this vision paper, we demonstrated how BLOGIC has its foundations in Peirce's existential graphs and provided a gentle introduction to the calculus of diagrammatic logic. Our contribution provides the translation of BLOGIC into a concrete RDF syntax, a semantic with FOL expressivity and multiple implementations. RDF Surfaces, with its N3S syntax, deviates from Hayes' original "annotated Turtle" and is a sub-language of Notation3 with a reduced set of features: N3 list terms to represent a set of graffiti blank nodes and graph terms to represent a (possibly empty) set of triples. Using these extensions to RDF and adding a surface type as a predicate, the full expressivity of FOL is available using simple entailment. The applicability of RDF Surfaces was demonstrated in two use cases, one from scholarly communication and one from healthcare. 

The benefits of RDF Surfaces as Web logic becomes evident in decentralized networks that need to exchange data and the logic behind the data. In our scholarly communication use-case, no facts were published by the researcher and institute, but conditions on future facts in the form of preferences. In the healthcare use case, the conditions on which medication would be applicable for (future) a patient were exchanged. In a decentralized network, Web agents should agree on a Web logic to analyze these preferences and conditions for possible triples that can become facts. 

Another use case that could benefit from RDF Surfaces is rights policies that need to exchange information about the permissions and prohibitions to access and use information. With RDF Surfaces, policy documents could not only make this information machine actionable but, in addition, in a decentralized setting, contradictions can be discovered between policies at the authoring time (and not at run time). The computer can say "no" to a policy author who does not have to wait for "the customer at the door" to find out something is wrong.

Publishing negative information is part of human activity; it is a part of commercial activity. The soft drink industry's "This drink contains no sugar" marketing statement is an explicit negative fact. For a consumer, this explicit negative information could be the reason to buy this drink. One option could be to add many negative predicates or classes to Web ontologies to publish this negative information. We believe this would obfuscate that classical negation is intended but not expressed and only adds to Hayes' "diamond of confusion."

Computation complexity is an issue, and as Web logic can be used for different purposes, many levels of complexity can be imagined. Adding the undecidability of FOL queries can potentially require vast computer resources or even run forever without providing an answer. Using Sowa argumentation~\cite{sowa_fads_2007}, we do not assume that worst-case scenarios will be the norm on the Web. The theorems of  Whitehead and Russell's \textit{Principia Mathematica} could already be proven in the 1960s by vacuum-tube machines. The Web runs today with a potential halting problem waiting in every HTML + JavaScript page. Database queries use the SQL language, which has FOL expressivity and can be solved in polynomial time. We advocate with RDF Surfaces to choose for sharing information and the logic behind this information using an RDF syntax, which makes as few assumptions as possible to create FOL expressivity. With FOL, we have semantics that is well understood.

Our research is still young and requires follow-up research to strengthen our claims. Formal semantics needs to be written. Implementations must demonstrate that values can be added to the Semantic Web using FOL expressivity for real-world use cases. Machines must cooperate and demonstrate, using proofs, why particular derivations were made and their reasoning for providing a conclusion. As Web citizens, we can clearly state what is and what is not using RDF Surfaces. 

\section{Acknowledgements}\label{s10}

This work is partly funded by the Andrew W. Mellon Foundation (grant number: 1903-06675), SolidLab Vlaanderen (Flemish Government, EWI and RRF project VV023/10), and the FWO Project FRACTION (Nr. G086822N). The authors would like to thank the W3C RDF Surfaces community group for joint discussions on the requirements for Web logic. The authors would also like to thank Jesse Wright and Ieben Smesseart for creating a Web version of the EYE reasoner and RDF Surfaces as a Web application. 



%
%
%
%
\nocite{*}
\bibliographystyle{ios1}           
\bibliography{paper.bib}        


\appendix

\renewcommand{\thesection}{Appendix \Alph{section}}  \renewcommand{\thesubsection}{\Alph{section}.\arabic{subsection}}
 
\section{Healthcare policies as RDF Surfaces in Notation3}

\begin{lstlisting}[numbers=none]
@prefix : <https://example.org/ns#> .
@prefix log: <http://www.w3.org/2000/10/swap/log#> .

(_:WHO) log:onNegativeSurface {

    _:WHO :has :AcuteMyocardialInfarction .
    
    () log:onNegativeSurface {
        _:WHO :has :AllergyForAspirin .
    } .
    
    () log:onNegativeSurface {
        _:WHO :has :ActivePepticUlcerDisease .
    } . 
    
    () log:onNegativeSurface {
        _:WHO :isPrescribed :aspirinLowDose .
    } .
    
} .

(_:WHO) log:onNegativeSurface {

    _:WHO :has :AcuteMyocardialInfarction .
    
    () log:onNegativeSurface {
        _:WHO :has :SevereAsthma .
    } .
    
    () log:onNegativeSurface {
        _:WHO :has :ChronicObstructivePulmonaryDisease .
    } .
    
    () log:onNegativeSurface {
        _:WHO :isPrescribed :betaBlocker .
    } .
} .
\end{lstlisting}

\section{Patient Ann data as RDF Surfaces in Notation3}

\begin{lstlisting}[numbers=none]
@prefix : <https://example.org/ns#> .
@prefix log: <http://www.w3.org/2000/10/swap/log#> .

# patient Ann
:Ann :has :Fever .

() log:onNegativeSurface {
    :Ann :has :AllergyForAspirin .
}.

() log:onNegativeSurface {
    :Ann :has :ActivePepticUlcerDisease .
} .
\end{lstlisting}

\section{Patient Joe data as RDF Surfaces in Notation3}

\begin{lstlisting}[numbers=none]
@prefix : <https://example.org/ns#> .
@prefix log: <http://www.w3.org/2000/10/swap/log#> .

# patient Joe
:Joe :has :AcuteMyocardialInfarction.
:Joe :has :AllergyForAspirin.

() log:onNegativeSurface {
    :Joe :has :ActivePepticUlcerDisease .
} .

() log:onNegativeSurface {
    :Joe :has :SevereAsthma .
} .

() log:onNegativeSurface {
    :Joe :has :ChronicObstructivePulmonaryDisease .
} .
\end{lstlisting}

\end{document}